\documentclass[twoside,12pt]{article}
\usepackage{epsf}
\let\ssection=\section
\renewcommand{\section}{\setcounter{equation}{0}\ssection}
\def\head#1#2{
\markboth{}{}
\setcounter{page}{#1}
\setcounter{section}{0}
\begin{huge}
\begin{flushleft}
\noindent\hangindent\parindent
{#2}
\end{flushleft}
\end{huge}
\bigskip
}
\def\auaddr#1#2{
{\noindent\LARGE\em#1}
{\noindent #2}
\medskip}

\def\and{
{\LARGE\em \&}
\bigskip
}

\def\refce{\small\smallskip\noindent\hangindent\parindent}

\def\caption#1{{ \begin{quote} \rm #1 \end{quote} }}

\begin{document}
\newcommand{\set}{{\tilde{D}}}
\newcommand{\all}{{D}}
\newcommand{\trace}{~{\large\rm Trace}~}
\newcommand{\s}{\varsigma}
\newcommand{\f}{\varphi}
\newcommand{\room}{\rule[-0.3cm]{0cm}{0.8cm}}
\newcommand{\smallroom}{\rule[-0.2cm]{0cm}{0.6cm}}
\newcommand{\hsp}{\hspace*{3mm}}
\newcommand{\vsp}{\vspace*{3mm}}
\newcommand{\nsp}{\vspace*{-3mm}}
\newcommand{\be}{\begin{equation}}
\newcommand{\ee}{\end{equation}}
\newcommand{\bd}{\begin{displaymath}}
\newcommand{\ed}{\end{displaymath}}
\newcommand{\bdm}{\begin{displaymath}}
\newcommand{\edm}{\end{displaymath}}
\newcommand{\bea}{\begin{eqnarray}}
\newcommand{\eea}{\end{eqnarray}}
\newcommand{\sgn}{~{\rm sgn}}
\newcommand{\extr}{~{\rm extr}}
\newcommand{\Equiv}{\Longleftrightarrow}
\newcommand{\pprime}{{\prime\prime}}
\newcommand{\notexists}{\exists\hspace*{-2mm}/}
\newcommand{\bra}{\langle}
\newcommand{\ket}{\rangle}
\newcommand{\bigbra}{\left\langle\room}
\newcommand{\bigket}{\right\rangle\room}
\newcommand{\bras}{\langle\!\langle}
\newcommand{\kets}{\rangle\!\rangle}
\newcommand{\bigbras}{\left\langle\!\!\!\left\langle\room}
\newcommand{\bigkets}{\room\right\rangle\!\!\!\right\rangle}
\newcommand{\order}{{\cal O}}
\newcommand{\minus}{\!-\!}
\newcommand{\plus}{\!+\!}
\newcommand{\erf}{{\rm erf}}
\newcommand{\bbf}{\mbox{\boldmath $f$}}
\newcommand{\bk}{\mbox{\boldmath $k$}}
\newcommand{\bm}{\mbox{\boldmath $m$}}
\newcommand{\br}{\mbox{\boldmath $r$}}
\newcommand{\bq}{\mbox{\boldmath $q$}}
\newcommand{\bu}{\mbox{\boldmath $u$}}
\newcommand{\bx}{\mbox{\boldmath $x$}}
\newcommand{\bz}{\mbox{\boldmath $z$}}
\newcommand{\bA}{\mbox{\boldmath $A$}}
\newcommand{\bB}{\mbox{\boldmath $B$}}
\newcommand{\bC}{\mbox{\boldmath $C$}}
\newcommand{\bF}{\mbox{\boldmath $F$}}
\newcommand{\bH}{\mbox{\boldmath $H$}}
\newcommand{\bJ}{\mbox{\boldmath $J$}}
\newcommand{\bK}{\mbox{\boldmath $K$}}
\newcommand{\bM}{\mbox{\boldmath $M$}}
\newcommand{\bQ}{\mbox{\boldmath $Q$}}
\newcommand{\bR}{\mbox{\boldmath $R$}}
\newcommand{\bW}{\mbox{\boldmath $W$}}
\newcommand{\hmu}{\hat{\mu}}
\newcommand{\hf}{\hat{f}}
\newcommand{\hQ}{\hat{Q}}
\newcommand{\hR}{\hat{R}}
\newcommand{\hbf}{\hat{\mbox{\boldmath $f$}}}
\newcommand{\hbh}{\hat{\mbox{\boldmath $h$}}}
\newcommand{\hbm}{\hat{\mbox{\boldmath $m$}}}
\newcommand{\hbr}{\hat{\mbox{\boldmath $r$}}}
\newcommand{\hbq}{\hat{\mbox{\boldmath $q$}}}
\newcommand{\hbD}{\hat{\mbox{\boldmath $D$}}}
\newcommand{\hbJ}{\hat{\mbox{\boldmath $J$}}}
\newcommand{\hbQ}{\hat{\mbox{\boldmath $Q$}}}
\newcommand{\hbR}{\hat{\mbox{\boldmath $R$}}}
\newcommand{\hbW}{\hat{\mbox{\boldmath $W$}}}
\newcommand{\bsigma}{\mbox{\boldmath $\sigma$}}
\newcommand{\btau}{\mbox{\boldmath $\tau$}}
\newcommand{\bomega}{\mbox{\boldmath $\Omega$}}
\newcommand{\bOmega}{\mbox{\boldmath $\Omega$}}
\newcommand{\bphi}{\mbox{\boldmath $\Phi$}}
\newcommand{\bpsi}{\mbox{\boldmath $\psi$}}
\newcommand{\bdelta}{\mbox{\boldmath $\Delta$}}
\newcommand{\btheta}{\mbox{\boldmath $\theta$}}
\newcommand{\bxi}{\mbox{\boldmath $\xi$}}
\newcommand{\bmu}{\mbox{\boldmath $\mu$}}
\newcommand{\brho}{\mbox{\boldmath $\rho$}}
\newcommand{\bEta}{\mbox{\boldmath $\eta$}}
\newcommand{\G}{{\cal G}}
\newcommand{\A}{{\cal A}}
\newcommand{\C}{{\cal C}}
\newcommand{\K}{{\cal K}}
\newcommand{\cA}{{\cal A}}
\newcommand{\cC}{{\cal C}}
\newcommand{\cD}{{\cal D}}
\newcommand{\cE}{{\cal E}}
\newcommand{\cF}{{\cal F}}
\newcommand{\cL}{{\cal L}}
\newcommand{\unity}{{\bf 1}\hspace{-1mm}{\bf I}}
\newcommand{\inn}{\!\cdot\!}
\newcommand{\LG}{\mbox{\normalsize ${\cal G}$}}
\newcommand{\LPsi}{\mbox{\normalsize $\Psi$}}
\newcommand{\hatildeG}{\bar{G}}
\newcommand{\ketset}{\mbox{\tiny ${\tilde{D}}$}}
\newcommand{\ketall}{\mbox{\tiny $D$}}
\newcommand{\ketop}{\mbox{\tiny ${\rm Q\!R\!P}$}}
\newcommand{\bsomega}{\!\!\mbox{\scriptsize\boldmath $\Omega$}}
\newcommand{\rmsets}{\!\!\mbox{\scriptsize\boldmath $\Xi$}}
\thispagestyle{plain}
\def\tit{Dynamics of Supervised Learning with Restricted
Training Sets}
\def\auth{A.C.C. Coolen $^\dag$ and D. Saad $^\ddag$}
\head{1}%{31}
{\tit}{
\auaddr{\auth}{\small
\\[2mm]
$\dag$ Department of Mathematics, King's College,
University of London\\
Strand, London WC2R 2LS, U.K.\\[2mm]
$\ddag$ Department of Computer Science and Applied
Mathematics,
Aston University\\
Aston Triangle, Birmingham B4 7ET, U.K.}
\markboth{A.C.C. Coolen and D. Saad}{\tit}
\def\ve{\varepsilon}
\begin{abstract}
\noindent
We study the dynamics of supervised learning in layered
neural
networks, in the regime where the size $p$ of the training
set is
proportional to the number $N$ of inputs. Here the local
fields are no
longer described by Gaussian probability distributions.  We
show how
dynamical replica theory can be used to predict the evolution
of macroscopic observables, including the relevant performance
measures, incorporating the old formalism in the limit
$\alpha=p/N\to\infty$
as a special case.  For simplicity we restrict ourselves to
single-layer networks and realizable tasks.
\end{abstract}
{\small
\baselineskip=4mm
\tableofcontents
}
\vfill
\pagebreak

\section{Introduction}

In the last few years much progress has been made in the
analysis of the dynamics of supervised learning in layered
neural
networks, using the strategy of statistical mechanics: by
deriving from the microscopic dynamical equations a set of
closed
laws describing the evolution of suitably chosen
macroscopic observables (dynamic order parameters) in the
limit of an
infinite system size\\[0mm]
[e.g. Kinzel and Rujan (1990), Kinouchi and
Caticha (1992), Biehl and Schwarze (1992,1995), Saad and
Solla (1995)].
A recent review and more extensive guide to the relevant
references
can be found in Mace and Coolen (1998a).  The main successful
procedure developed so far is built on the following
cornerstones:
\vspace*{-2.5mm}
\begin{itemize}
\item {\em The task to be learned is defined by a (possibly
noisy)
`teacher', which is itself a layered neural network.}  This
induces a
canonical set of dynamical order parameters, typically the
(rescaled)
overlaps between the various student weight vectors and the
corresponding teacher weight vectors.
\vspace*{-2.5mm}
\item {\em The number of network inputs is (eventually) taken
to be
infinitely large.}  This ensures that fluctuations in
mean-field
observables will vanish and creates the possibility of using
the central limit theorem.
\vspace*{-2.5mm}
\item {\em The number of `hidden' neurons is finite.}  This
prevents
the number of order parameters from being infinite, and
ensures that
the cumulative impact of their fluctuations is insignificant.
\vspace*{-2.5mm}
\item {\em The size of the training set is much larger than
the number
of updates made.}  Each example presented is now different
from those that have already been seen, such that the local
fields
will have Gaussian probability distributions, which leads to
closure of the
dynamic equations.
\end{itemize}
\vspace*{-3mm}

\noindent
These are not ingredients to simplify the calculations, but
vital
conditions, without which the standard method fails. Although
the
assumption of an infinite system size has been shown not to be
 too critical (Barber et al, 1996), the
other assumptions do
place serious restrictions on the degree of realism of the
scenarios
that can be analyzed, and have thereby, to some extent,
prevented
the theoretical results from being used by practitioners.

In this paper we study the dynamics of learning in layered
neural
networks with restricted training sets, where the number $p$
of
examples (`questions' with corresponding `answers') scales
linearly
with the number $N$ of inputs, i.e. $p=\alpha N$.
Here individual questions will
re-appear during the learning process as soon as the number of
weight
updates made is of the order of the size of the training set.
In the
traditional models, where the duration of an update is
defined as
$N^{-1}$, this happens as soon as $t=\order(\alpha)$. At that
point
correlations develop between the weights and the questions in
the
training set, and the dynamics is of a spin-glass type, with
the
composition of the training set playing the role of `quenched
disorder'. The main consequence of this is that the central
limit
theorem no longer applies to the student's local fields,
which are now
described by non-Gaussian distributions.  To demonstrate this
we trained (on-line) a perceptron with weights $J_{i}$
on noiseless examples generated by a teacher perceptron with
weights $B_i$, using the Hebb and AdaTron rules.  We plotted
in Fig. 1
the student and teacher fields, $x = \bJ\inn\bxi$ and $y =
\bB\inn\bxi$ respectively, where $\bxi$ is the input vector,
for
$p=N/2$ examples and at time $t=50$. The marginal distribution
$P(x)$ for $p=N/4$, at times $t=10$ for the Hebb rule and
$t=20$ for the
Adatron rule, is shown in Fig.~2.  The non-Gaussian student
field
distributions observed in Figs.~1 and 2 induce a deviation
between the
training- and generalization errors, which measure the network
performance on training and test examples, respectively. The
former
involves averages over the non-Gaussian field distribution,
whereas
the latter (which is calculated over {\em all} possible
examples) still involves Gaussian fields.

The appearance of
non-Gaussian  fields leads to a  breakdown of the standard
formalism,  based on deriving closed equations for
a finite number of observables: the field distributions can no
longer be characterized by a few moments, and the macroscopic
laws must now be averaged over realizations of the training
set. One could still try to use Gaussian distributions as
large $\alpha$ approximations, see e.g.  Sollich and Barber
(1998), but it
will be clear from Figs. 1 and 2 that a systematic theory
will have to give up Gaussian distributions entirely. The
first rigorous study of the dynamics of learning with
restricted
training sets in non-linear networks, via the calculation of
generating functionals, was carried out by Horner (1992) for
perceptrons with binary weights.
In this paper we show how the formalism of dynamical replica
theory (see e.g. Coolen et al, 1996) can be used successfully
to
predict the evolution of macroscopic observables for finite
$\alpha$,
incorporating the infinite training set formalism
 as a special case, for $\alpha\to\infty$.  Central to our
approach is the
derivation of a diffusion equation for the joint distribution
of the
student and teacher fields, which will be found to have
Gaussian
solutions only for $\alpha\to\infty$.  For simplicity and
transparency
we restrict ourselves to single-layer systems and noise-free
teachers.
Application and generalization of our methods to multi-layer
systems
(Saad and Coolen, 1998) and learning scenarios involving
`noisy'
teachers (Mace and Coolen, 1998b) are presently under way.

This paper is organized as follows. In section 2 we first
derive a
Fokker-Planck equation describing the evolution of arbitrary
mean-field observables for $N\to\infty$. This allows us to
identify the conditions for the latter to be described by
closed
deterministic laws. In section 3 we choose as our observables
the
joint field distribution $P[x,y]$, in addition to (the
traditional ones) $Q$ and $R$, and show that this set obeys
deterministic laws.  In order to close these laws
we use the tools of dynamical replica theory.  Details of the
replica calculation are given in section 4, to be skipped by
those primarily interested in results.  In section 5 we show
how in
the limit $\alpha\to\infty$ (infinite training sets) the
equations of
the conventional theory are recovered. Finally we work out
our equations explicitly for the example of on-line Hebbian
learning with restricted training sets, and compare our
predictions with exact results (derived directly from the
microscopic
equations) and with numerical simulations.  \vfill

\begin{figure}[t]
\vspace*{75mm}
\hbox to \hsize{\hspace*{-0cm}\includegraphics{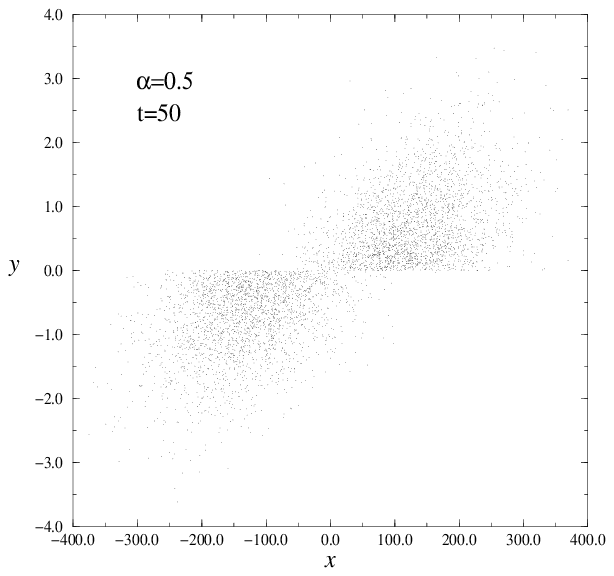}
\hspace*{0cm}}
\vspace*{-4.8mm}
\hbox to
\hsize{\hspace*{62mm}\includegraphics{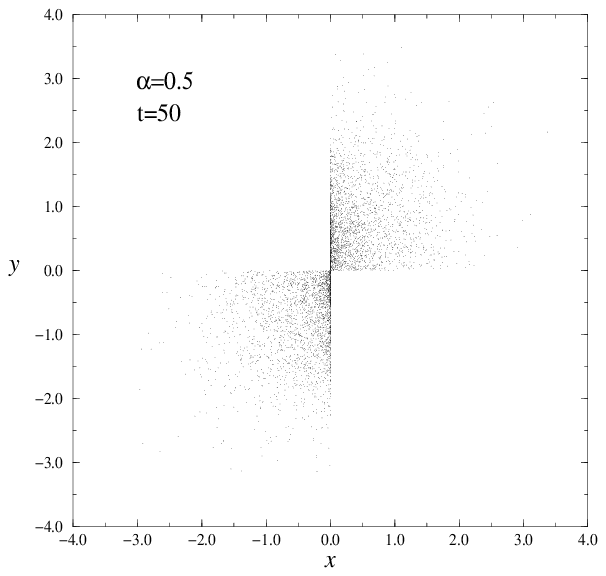}
\hspace*{-65mm}}
\vspace*{-6mm}
\caption{Fig. 1: Student and teacher fields $(x,y)$ as
observed during
numerical simulations of on-line learning (learning rate
$\eta=1$) in
a perceptron of size $N=10,000$ at $t=50$, using `questions'
from a
restricted training set of size $p=\frac{1}{2} N$.  Left:
Hebbian
learning. Right: AdaTron learning.  Note: in the case of
Gaussian
field distributions one would have found spherically shaped
plots.}
\vspace*{72mm}
\hbox to \hsize{\hspace*{-0cm}\includegraphics{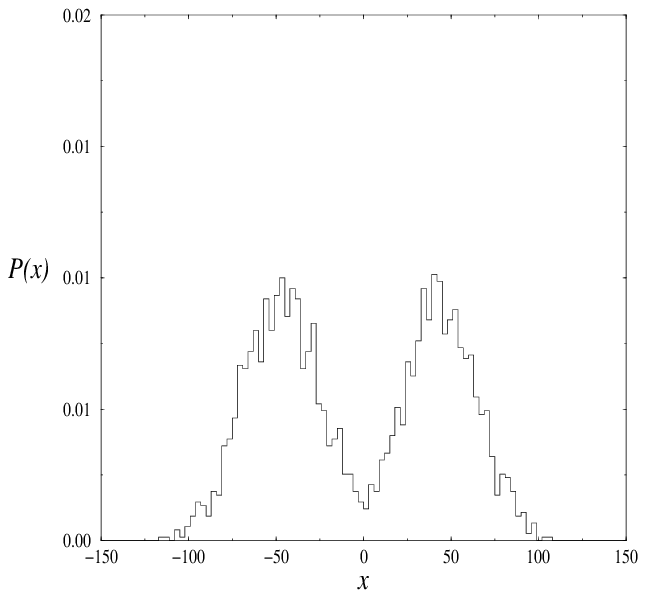}
\hspace*{0cm}}
\vspace*{-4.8mm}
\hbox to
\hsize{\hspace*{62mm}\includegraphics{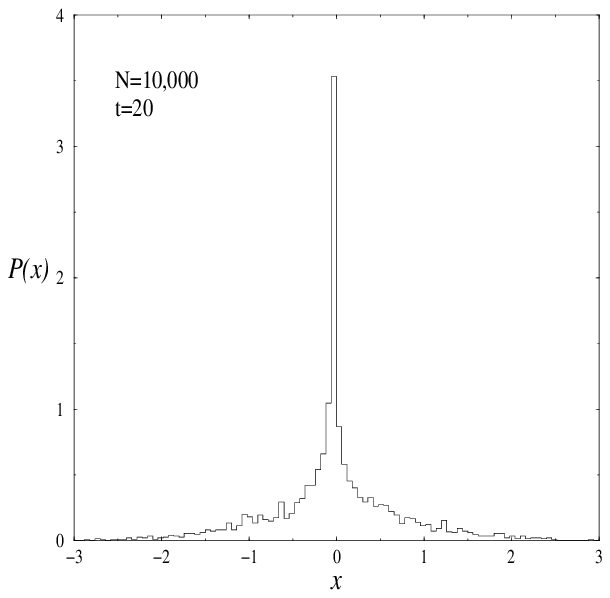}
\hspace*{-62mm}}
\vspace*{-9mm}
\caption{Fig. 2: Distribution $P(x)$ of student fields as
observed
during numerical simulations of on-line learning (learning
rate
$\eta=1$) in a perceptron of size $N=10,000$, using
`questions' from a restricted training set of size
$p=\frac{1}{4} N$.
Left: Hebbian learning, measured at $t=10$.  Right: AdaTron
learning,
measured at $t=20$.  Note: not only are these distributions
distinctively non-Gaussian, they also appear to vary widely
in their
basic characteristics, depending on the learning rule used.}
\end{figure}
\clearpage

%%%%%%%%%%%%%%%%%%%%%%%%%%%%%%%%%%%%%%%%%%%%%%%%%%%%%%%%

\section{From Microscopic to Macroscopic Laws}
\subsection{Definitions}

A student perceptron operates the
following rule, which is parametrised by the weight vector
$\bJ\in\Re^N$:
\be
S:\{-1,1\}^N\to\{-1,1\} ~~~~~~~~~~
S(\bxi)=\sgn\left[\bJ\cdot\bxi\right]
\label{eq:student}
\ee
It tries to emulate the operation of a teacher perceptron,
via an
iterative procedure for updating its parameters $\bJ$. The
teacher
perceptron operates a similar rule, characterized by a given
(fixed)
weight vector $\bB\in\Re^N$:
\be
T:\{-1,1\}^N\to\{-1,1\} ~~~~~~~~~~
T(\bxi)=\sgn\left[\bB\cdot\bxi\right]
\label{eq:teacher}
\ee
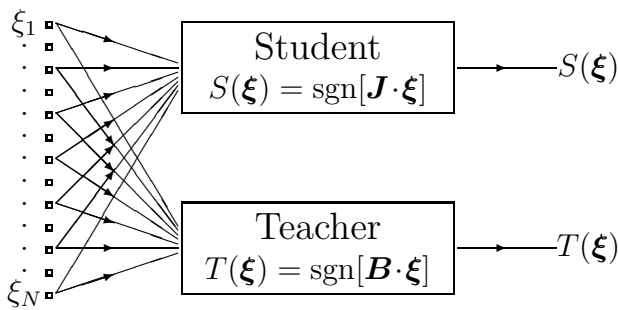
\begin{figure}[h]
\setlength{\unitlength}{0.06mm}
\newcommand{\sq}{\thinlines\framebox(10,10)}
\newcommand{\here}{\makebox(0,0)}
\begin{picture}(1300,750)(-100,350)
\newcommand{\Smallsize}{130}
\newcommand{\Size}{270}
\put(50,1000){\here{$\xi_1$}}
\put(50,950){\here{$.$}}
\put(50,900){\here{$.$}}
\put(50,850){\here{$.$}}
\put(50,800){\here{$.$}}
\put(50,750){\here{$.$}}
\put(50,700){\here{$.$}}
\put(50,650){\here{$.$}}
\put(50,600){\here{$.$}}
\multiput(100,390)(0,50){13}{\sq}
\put(50,550){\here{$.$}}
\put(50,500){\here{$.$}}
\put(50,450){\here{$.$}}
\put(50,400){\here{$\xi_N$}}
\put(400,800){\framebox(600,200)}
\put(700,950){\here{\large Student}}
\put(700,850){\here{$S(\bxi)={\rm sgn}[\bJ\inn\bxi]$}}
\put(1010,900){\line(1,0){220}}
\put(1010,900){\vector(1,0){110}}
\put(1300,900){\here{$S(\bxi)$}}
\put(120,400){\line(3,1){\Size}}
\put(120,500){\line(3,4){\Size}}
\put(120,600){\line(1,1){\Size}}
\put(120,700){\line(3,2){\Size}}
\put(120,800){\line(3,1){\Size}}
\put(120,900){\line(1,0){\Size}}
\put(120,1000){\line(3,-1){\Size}}
\put(120,400){\line(3,5){\Size}}
\put(120,500){\line(1,0){\Size}}
\put(120,600){\line(3,-1){\Size}}
\put(120,700){\line(3,-2){\Size}}
\put(120,800){\line(1,-1){\Size}}
\put(120,900){\line(3,-4){\Size}}
\put(120,1000){\line(3,-5){\Size}}
\put(400,400){\framebox(600,200)}
\put(700,550){\here{\large Teacher}}
\put(700,450){\here{$T(\bxi)={\rm sgn}[\bB\inn\bxi]$}}
\put(1010,500){\line(1,0){220}}
\put(1010,500){\vector(1,0){110}}
\put(1300,500){\here{$T(\bxi)$}}
\put(120,400){\vector(3,1){\Smallsize}}
\put(120,500){\vector(3,4){\Smallsize}}
\put(120,600){\vector(1,1){\Smallsize}}
\put(120,700){\vector(3,2){\Smallsize}}
\put(120,800){\vector(3,1){\Smallsize}}
\put(120,900){\vector(1,0){\Smallsize}}
\put(120,1000){\vector(3,-1){\Smallsize}}
\put(120,500){\vector(1,0){\Smallsize}}
\put(120,600){\vector(3,-1){\Smallsize}}
\put(120,700){\vector(3,-2){\Smallsize}}
\put(120,800){\vector(1,-1){\Smallsize}}
\put(120,900){\vector(3,-4){\Smallsize}}
\end{picture}
\caption{Fig. 3: Supervised learning in perceptrons.}
\label{fig:teacherstudent}
\end{figure}
In order to do so, the student perceptron modifies its
weight
vector
$\bJ$ according to an iterative procedure, using examples
of
input
vectors (or `questions') $\bxi$, drawn at random from a
fixed
training
set $\set\subseteq \all=\{-1,1\}^N$, and the corresponding
values of
the teacher outputs $T(\bxi)$, see Fig. 3.

We consider the case where the training set is a randomly
composed
subset $\set\subset D$, of size $|\set|=p=\alpha N$ with
$\alpha>0$:
\be \set=\{\bxi^1,\ldots,\bxi^p\}~~~~~~~~~~p=\alpha N
~~~~~~~~~~
\bxi^\mu\in\all~~{\rm for~all}~\mu
\label{eq:trainingset}
\ee
We will denote averages over the training set $\set$ and
averages
over the full question set $D$ in the following way:
\bd \bra
\Phi(\bxi) \ket_{\ketset}=\frac{1}{|\set|}\sum_{\bxi\in\set}
\Phi(\bxi)
~~~~~{\rm and}~~~~~~ \bra \Phi(\bxi)\ket_{\ketall}=
\frac{1}{|\all|}\sum_{\bxi\in
\all}\Phi(\bxi) \ .  \ed
We will analyze the following two classes of learning rules:
\be
\begin{array}{ll}
{\rm on\!-\!line:} &
\bJ(m\plus 1)=
\bJ(m)+
\frac{\eta}{N} \
\bxi(m) \ \G\left[\bJ(m)\inn\bxi(m), \bB\inn\bxi(m)\right]
\\[2mm]
{\rm batch:} &
\bJ(m\plus 1)=
\bJ(m)+
\frac{\eta}{N} \
\bra \bxi~ \G\left[\bJ(m)\inn\bxi, \bB\inn\bxi\right]
\ket_{\ketset}
\end{array}
\label{eq:weightdynamics}
\ee
In on-line learning one draws at each iteration step $m$ a
question
$\bxi(m)\in\set$ at random, the dynamics is thus a stochastic
process;
in batch learning one iterates a deterministic map.  The
function
$\G[x,y]$ is assumed to be bounded and not to depend on $N$,
other
than via its two arguments.

Our most important observables during learning are the
training error
$E_{\rm t}(\bJ)$ and the generalization error
$E_{\rm g}(\bJ)$, defined as follows:
\be E_{\rm t}(\bJ)=\bra \theta[-(\bJ\inn\bxi) (\bB\inn\bxi)]
\ket_{\ketset}  \ ,
\smallroom
\label{eq:Et}
\ee
\be
E_{\rm g}(\bJ)=\bra \theta[-(\bJ\inn\bxi) (\bB\inn\bxi)]
\ket_{\ketall} \ .
\smallroom
\label{eq:Eg}
\ee
Only if the training set $\set$ is sufficiently large, and if
there are no correlations between $\bJ$ and the questions
$\bxi\in\set$, will these two errors will be identical.

\subsection{From Discrete to Continuous Time}

We next convert the dynamical laws (\ref{eq:weightdynamics})
into the
language of stochastic processes. We introduce the probability
$\hat{p}_m(\bJ)$ to find weight vector $\bJ$ at discrete
iteration
step $m$. In terms of this microscopic probability
distribution the
processes (\ref{eq:weightdynamics}) can be written in the
general Markovian form
\be \hat{p}_{m+1}(\bJ)=
\int\!d\bJ^\prime~W[\bJ;\bJ^\prime] \ \hat{p}_m(\bJ^\prime)
\ ,
\label{eq:markovprocess}
\ee
with the transition probabilities
\be
\begin{array}{ll}
{\rm on\!-\!line:} &
W[\bJ;\bJ^\prime]=\bra \delta\left[
\bJ\minus \bJ^\prime\minus
\frac{\eta}{N} \ \bxi \ \G\left[\bJ^\prime\inn\bxi,
\bB\inn\bxi\right]
\right]\ket_{\ketset}
\\[2mm]
{\rm batch:} &
W[\bJ;\bJ^\prime]=
\delta\left[\bJ\minus \bJ^\prime\minus
\frac{\eta}{N}\bra \bxi~\G\left[\bJ^\prime\inn\bxi,
\bB\inn\bxi\right]\ket_{\ketset}
\right]
\end{array}
\label{eq:transitionmatrix}
\ee
We now make the transition to a description involving
real-valued time
labels by choosing the duration of each iteration step to be
a real-valued random number, such that the probability that
at time $t$
precisely $m$ steps have been made is given by the Poisson
expression
\be \pi_m(t)=\frac{1}{m!}(Nt)^me^{-Nt} \ .
\label{eq:poisson}
\ee
For times $t\gg N^{-1}$ we find
$t=m/N+\order(N^{-\frac{1}{2}})$,
the usual time unit.  Due to the random durations of the
iteration
steps we have to switch to the following microscopic
probability
distribution: \be p_t(\bJ)= \sum_{m\geq 0}\pi_m(t) \
\hat{p}_m(\bJ) \
.
\label{eq:conttime}
\ee
This distribution obeys a simple differential equation, which
immediately follows from the pleasant properties of
(\ref{eq:poisson})
under temporal differentiation:
\be \frac{d}{dt} \ p_t(\bJ) =
N\int\!d\bJ^\prime~\left\{ W[\bJ;\bJ^\prime]-\delta[
\bJ\minus\bJ^\prime]\right\} \ p_t(\bJ^\prime) \ .
\label{eq:conttimemarkov}
\ee
So far no approximations have been made, equation
(\ref{eq:conttimemarkov}) is exact for any $N$. It is the
equivalent
of the master equation often introduced to define the
dynamics of spin
systems.

\subsection{Derivation of Macroscopic Fokker-Planck Equation}

We now wish to investigate the dynamics of a number of as yet
arbitrary {\em macroscopic} observables
$\bomega[\bJ]=(\Omega_1[\bJ],\ldots,\Omega_k[\bJ])$. To do
so we
introduce a macroscopic probability distribution
\be P_t(\bOmega)=
\int\!d\bJ~p_t(\bJ) \ \delta\left[\bOmega-\bOmega[\bJ]\right] \ .
\ee
Its time derivative immediately follows from that in
(\ref{eq:conttimemarkov}):
\be
\frac{d}{dt}P_t(\bOmega) =
\int\!d\bOmega^\prime~ {\cal W}_t[\bOmega;\bOmega^\prime] \
P_t(\bOmega^\prime) \ ,
\label{eq:macrodynamics}
\ee
where
\bd
{\cal W}_t[\bOmega;\bOmega^\prime]\! =\!
\frac{\int\!d\bJ^\prime~p_t(\bJ^\prime) \
\delta\left[\bOmega^\prime\minus \bOmega[\bJ^\prime]\right]
\
\int\!d\bJ \ \delta\left[\bOmega\minus \bOmega[\bJ]\right]
\! N \!
\left\{ W[\bJ;\bJ^\prime]\minus \delta[\bJ\minus \bJ^\prime]
\right\} }
{\int\!d\bJ^\prime~p_t(\bJ^\prime) \ \delta
\left[\bOmega^\prime\minus
\bOmega[\bJ^\prime]\right]}
\ed
If we insert the relevant expressions
(\ref{eq:transitionmatrix}) for
$W[\bJ;\bJ^\prime]$ we can perform the $\bJ$-integrations,
and obtain
expressions in terms of so-called sub-shell averages,
defined as \bd
\bra f(\bJ)\ket_{ \ \bsomega;t}= \frac{\int\!d\bJ~p_t(\bJ) \
\delta\left[\bOmega\minus \bOmega[\bJ]\right] \ f(\bJ)}
{\int\!d\bJ~p_t(\bJ) \ \delta\left[\bOmega\minus \bOmega
[\bJ]\right]}
\ .
\ed
For the two types of learning rules at hand we obtain:
\bd {\cal
W}^{\rm onl}_t[\bOmega;\bOmega^\prime]=N \bigbras \delta
\left[
\bOmega\minus
\bOmega[\bJ\plus\frac{\eta}{N} \ \bxi \
\G[\bJ\cdot\bxi,\bB\cdot\bxi]]
\right]\bigket_{\!\ketset}  \minus \delta
\left[\bOmega\minus \bOmega[\bJ]\right]
\bigket_{\bsomega^\prime;t} \ed \bd {\cal W}^{\rm
bat}_t[\bOmega;\bOmega^\prime] = N\bigbra \delta
\left[\bOmega\minus
\bOmega[\bJ
\plus\frac{\eta}{N}\bra\bxi \ \G[\bJ\cdot\bxi,\bB\cdot\bxi]
\ket_\Omega]
\right] \minus \delta\left[\bOmega\minus \bOmega[\bJ]\right]
\bigket_{\bsomega^\prime;t}
\ed
We now insert integral representations for the
$\delta$-distributions.
This gives for our two learning scenario's:
\be {\cal W}^{\rm onl}_t[\bOmega;\bOmega^\prime]\!=\!
\int\!\frac{d\hat{\bOmega}}{(2\pi)^k}e^{i\hat{\bOmega}
\cdot\bOmega}~
N\bigbras e^{-i\hat{\bOmega}\cdot \bOmega\left[\bJ
\plus\frac{\eta}{N}\bxi\LG\left[\bJ\cdot\bxi,\bB\cdot\bxi
\right]\right]} \bigket_{\!\!\ketset}
\minus e^{-i\hat{\bOmega}\cdot\bOmega\left[\bJ\right]}
\bigket_{\bsomega^\prime;t}
\label{eq:macromap1}
\ee
\be
{\cal W}^{\rm bat}_t[\bOmega;\bOmega^\prime]\!=\!
\int\!\frac{d\hat{\bOmega}}{(2\pi)^k}e^{i\hat{\bOmega}
\cdot\bOmega}~
N\bigbra
e^{-i\hat{\bOmega}\cdot\bOmega
\left[\bJ
\plus\frac{\eta}{N}\bigbra\bxi\LG\left[\bJ\cdot\bxi,
\bB\cdot\bxi\right]\bigket_{\!\!\ketset} \right]}
\minus
e^{-i\hat{\bOmega}\cdot\bOmega\left[\bJ\right]}
\bigket_{\bsomega^\prime;t}
\label{eq:macromap2}
\ee
Still no approximations have been made. The above two
expressions
differ only in the stage where the averaging over the
training set
is carried out.

In expanding equations (\ref{eq:macromap1},\ref{eq:macromap2})
for large $N$ and finite $t$ we have to be careful, since the
system size
$N$ enters both as a small parameter to control the magnitude
of the modification of individual components of the weight
vector, but also
determines the dimensions and lengths of various vectors that
occur.
We therefore inspect more closely the usual Taylor expansions:
\bd F[\bJ\plus\bk] =\sum_{\ell\geq 0}\frac{1}{\ell
!}\sum_{i_1=1}^N\cdots\sum_{i_\ell=1}^N k_{i_1}
\cdots k_{i_\ell}
\frac{\partial^\ell F[\bJ]}{\partial J_{i_1}\cdots
\partial J_{i_\ell}} \ .
\ed
If all derivatives were to be treated as $\order(1)$ (i.e.
if we only
take into account the scaling of the shift $\bk$ with $N$),
problems
could arise, since in the cases of interest (where
$\bk^2=\order(N^{-1})$) this series could diverge as
$F[\bJ\plus\bk]=
\sum_{\ell\geq 0}(\sum_i k_i)^\ell =\sum_{\ell\geq 0}
\order(1)$.  If
we assess how derivatives with respect to individual
components $J_i$
scale for the standard types of mean-field observables, we
find the
following scaling property which we will choose as our {\em
definition} of mean-field observables:
\be
\frac{\partial^\ell F[\bJ]}{\partial
J_{i_1}\cdots\partial J_{i_\ell}}=\order\left(\!
\frac{F[\bJ]}{|\bJ|^\ell} \ N^{\frac{1}{2}\ell-d}
\!\right)
~~~~~~(N\to\infty) \ ,
\label{eq:meanfield}
\ee
in which $d$ is the number of {\em different} elements in the
set $\{i_1,\ldots,i_\ell\}$.  If $F[\bJ]$ is a mean-field
observable in
the sense of (\ref{eq:meanfield}), we can estimate the
scaling of the
various terms in the Taylor expansion:
\be F[\bJ\plus\bk]= F[\bJ]+\sum_i k_i\frac{\partial
F[\bJ]}{\partial
J_i} +\frac{1}{2}\sum_{ij}k_i k_j
\frac{\partial^2 F[\bJ]}{\partial J_i
\partial J_j} +\sum_{\ell\geq 3} \order\left(\!
F[\bJ]\left[\frac{|\bk|}{|\bJ|}\right]^\ell\!\right)
\label{eq:taylor}
\ee
(in the last step we have used $\sum_i k_i=
\order(\sqrt{N}|\bk|)$).

We apply (\ref{eq:taylor}) to our macroscopic equations
(\ref{eq:macromap1},\ref{eq:macromap2}), restricting
ourselves from
now on to mean-field observables $\bOmega[\bJ]=\order(N^0)$
in the
sense of (\ref{eq:meanfield}), one of which we choose to be
$\bJ^2$. Here the shifts $\bk$, being either
$\frac{\eta}{N}\bxi \
\G[\bJ\cdot\bxi;\bB\cdot\bxi]$ or $\frac{\eta}{N}\bra\bxi \
\G[\bJ\cdot\bxi;\bB\cdot\bxi]\ket_{\ketset} $, scale as
$|\bk|=\order(N^{-\frac{1}{2}})$. Consequently the
$\ell$-th order
term in the expansions of both (\ref{eq:macromap1}) and
(\ref{eq:macromap2}) will be of order $N^{-\frac{\ell}{2}}$:
\bd
e^{-i\hat{\bOmega} \cdot \bOmega\left[\bJ\plus\bk\right]} =
e^{-i\hat{\bOmega} \cdot \bOmega\left[\bJ\right]}\left\{
1\minus i\sum_i
k_i\frac{\partial}{\partial J_i}(\hat{\bOmega}\cdot\bOmega
[\bJ])
\right.  ~~~~~~~~~~~~~~~~~~~~~~~~~~~~~~~~~~~~~~~~~
\ed
\bd \left.
- \frac{i}{2}\sum_{ij}k_i k_j\frac{\partial^2}{\partial J_i
\partial J_j} (\hat{\bOmega}\cdot\bOmega[\bJ]) \minus
\frac{1}{2}\left[ \sum_i k_i\frac{\partial}{\partial
J_i}(\hat{\bOmega}\cdot\bOmega[\bJ]) \right]^2\right\}
+\order(N^{-\frac{3}{2}}) \ .
\ed
This, in turn, gives
\bd
\int\!\frac{d\hat{\bOmega}}{(2\pi)^k}e^{i\hat{\bOmega} \
\cdot \ \bOmega}~
N\left[e^{-i\hat{\bOmega}  \cdot  \bOmega\left[\bJ\plus\bk
\right]} \minus
e^{-i\hat{\bOmega}  \cdot  \bOmega\left[\bJ\right]}\right]
~~~~~~~~~~~~~~~~~~~~~~~~~~~~~~~~~~~~~~~~~~~~~~~~~~~~~~~~~~~~~~~~~~~~~~
\ed \bd =-N \left\{ \sum_{\mu}\frac{\partial}{\partial
\Omega_\mu}
\left[ \sum_i k_i\frac{\partial \Omega_\mu[\bJ]}{\partial J_i}
\plus
\frac{1}{2}\sum_{ij}k_i k_j
\frac{\partial^2\Omega_\mu[\bJ]}{\partial
J_i \partial J_j}\right] \right. ~~~~~~~~~~~~~~~~~~~~~~~~
\ed
\bd
\left.  ~~~~~~~~~~~~~~
-\frac{1}{2}\sum_{\mu\nu}\frac{\partial^2}{\partial\Omega_\mu
\partial\Omega_\nu} \sum_{ij}
k_ik_j\frac{\partial\Omega_\mu[\bJ]}{\partial J_i}
\frac{\partial\Omega_\nu[\bJ]}{\partial J_j}\right\}
\delta\left[\bOmega-\bOmega[\bJ]\right] +
\order(N^{-\frac{1}{2}}) \ .
\ed
It is now evident, in view of
(\ref{eq:macromap1},\ref{eq:macromap2}),
that both types of dynamics are described by macroscopic laws
with transition probability densities of the general form
\bd {\cal
W}^{\rm\star\star\star}_t[\bOmega;\bOmega^\prime]= \left\{
-\sum_{\mu}F_\mu[\bOmega^\prime;t]\frac{\partial}{\partial
\Omega_\mu}
+ \frac{1}{2}\sum_{\mu\nu}G_{\mu\nu}[\bOmega^\prime;t]
\frac{\partial^2}{\partial\Omega_\mu\partial\Omega_\nu}
\right\}\delta\left[\bOmega\minus \bOmega^\prime\right]
\vspace*{-2mm}
\ed
\bd
~~~~~~~~~~~~~~~~~~~~~~~~~~~~~~~~~~~~~~~
~~~~~~~~~~~~~~~~~~~~~~~~~~~~~~~~~~~~~~~~~~~~~~~~
+~\order(N^{-\frac{1}{2}})
\ed
which, due to (\ref{eq:macrodynamics}) and for $N\to\infty$
and finite times,
leads to a Fokker-Planck equation:
\be
\frac{d}{dt}P_t(\bOmega)
=
-\sum_{\mu=1}^k\frac{\partial}{\partial\Omega_\mu}
\left\{F_\mu[\bOmega;t]P_t(\bOmega)\right\}
+\frac{1}{2}\sum_{\mu\nu=1}^k
\frac{\partial^2}{\partial\Omega_\mu\partial\Omega_\nu}
\left\{ G_{\mu\nu}[\bOmega;t]P_t(\bOmega)\right\} \ .
\label{eq:fokkerplanck}
\ee
The differences between the two types of dynamics are in
the explicit
expressions for the flow- and diffusion terms:
\bd
F^{\rm onl}_\mu[\bOmega;t]=
\eta \ \bigbras\sum_i \xi_i \ \G[\bJ\cdot\bxi,\bB\cdot\bxi]
\
\frac{\partial \Omega_\mu[\bJ]}{\partial
J_i}\bigket_{\!\!\ketset} \bigket_{\bsomega;t}
~~~~~~~~~~~~~~~~~~~~~~~~~~~~~~~~~~~~~~
\ed
\bd
~~~~~~~~~~~~~~~~~~~~~~~~~~~~~~
+ \frac{\eta^2}{2N}\bigbras\sum_{ij}
\xi_i\xi_j \ \G^2[\bJ\cdot\bxi,\bB\cdot\bxi] \
\frac{\partial^2\Omega_\mu[\bJ]}{\partial J_i \partial J_j}
\bigket_{\!\!\ketset} \bigket_{\bsomega;t}
\ed
\bd
G^{\rm onl}_{\mu\nu}[\bOmega;t]=
\frac{\eta^2}{N} \ \bigbras \sum_{ij}
\xi_i\xi_j \ \G^2[\bJ\cdot\bxi,\bB\cdot\bxi]
\left[\frac{\partial\Omega_\mu[\bJ]}{\partial J_i}\right]
\left[\frac{\partial\Omega_\nu[\bJ]}{\partial J_j}\right]
\bigket_{\!\!\ketset} \bigket_{\bsomega;t}
~~~~~~~~~~~~~~~~~~~~~~~~~~~~~~~
\ed
\bd
F^{\rm bat}_\mu[\bOmega;t]=
\eta \ \bigbra\sum_i
\bra\xi_i \ \G[\bJ\cdot\bxi,\bB\cdot\bxi] \ \ket_{\ketset}
\frac{\partial \Omega_\mu[\bJ]}{\partial J_i}
\bigket_{\bsomega;t}
~~~~~~~~~~~~~~~~~~~~~~~~~~~~~~~~~~~~~~~~~~~~~~
\ed
\bd
~~~~~~~~~~~~~~~~~~~~
+ \frac{\eta^2}{2N} \ \bigbra\sum_{ij}
\bra\xi_i \ \G[\bJ\cdot\bxi,\bB\cdot\bxi]
\ket_{\ketset}
\bra\xi_j \ \G[\bJ\cdot\bxi,\bB\cdot\bxi]
\ket_{\ketset}  \
\frac{\partial^2\Omega_\mu[\bJ]}{\partial J_i \partial J_j}
\bigket_{\bsomega;t}
\ed
\bd
G^{\rm bat}_{\mu\nu}[\bOmega;t] =
\frac{\eta^2}{N} \!\!\bigbra \! \sum_{ij}
\bra\xi_i\G[\bJ\!\cdot\!\bxi,\bB\!\cdot\!\bxi]
\ket_{\ketset}  \
\bra\xi_j\G[\bJ\!\cdot\!\bxi,\bB\!\cdot\!\bxi]
\ket_{\ketset}
\left[\frac{\partial\Omega_\mu[\bJ]}{\partial J_i}\right] \!
\left[\frac{\partial\Omega_\nu[\bJ]}{\partial J_j}\right]
\bigket_{\bsomega;t}
\ed
Equation (\ref{eq:fokkerplanck}) allows us to define the goal
of our
exercise in more explicit form. If we wish to arrive at closed
deterministic macroscopic equations, we have to choose our
observables
such that
\begin{enumerate}
\item $\lim_{N\to\infty}G_{\mu\nu}[\bOmega;t]=0$ ~~~~~~~~
(this ensures
determinism)
\item $\lim_{N\to\infty}\frac{\partial}{\partial
t}F_\mu[\bOmega;t]=0$ ~~~~~~~(this ensures closure)
\end{enumerate}
In the case of time-dependent global parameters, such as
learning
rates or decay rates, the latter condition relaxes to the
requirement
that any explicit time-dependence of $F_\mu[\bOmega;t]$ is
restricted to these global parameters.

\section{Application to Canonical Observables}

\subsection{Choice of Canonical Observables}

We now apply the general results obtained so far to a
specific set of
observables $\bOmega[\bJ]$, taylored to the problem at hand:
\be
Q[\bJ]=\bJ^2,~~~~~
R[\bJ]=\bJ\inn\bB,~~~~~
P[x,y;\bJ]=\bra\delta[x\minus \bJ\inn\bxi] \
\delta[x\minus\bB\inn\bxi]\ket_{\ketset}
\label{eq:learningobservables}
\ee
This choice is motivated by the following considerations:
(i) in order
to incorporate the standard theory in the limit
$\alpha\to\infty$ we
need at least $Q[\bJ]$ and $R[\bJ]$, (ii) we need to be
able to
calculate the training error, which involves field statistics
calculated over the training set $\set$, as described by
$P[x,y;\bJ]$,
and (iii) for finite $\alpha$ one cannot expect closed
macroscopic
equations for just a finite number of order parameters,
the present
choice (involving the order parameter {\em function}
$P[x,y;\bJ]$)
represents effectively an infinite number\footnote{A
simple rule of
thumb is the following: if a process requires replica theory
for its
stationary state analysis, as does learning with restricted
training
sets, its dynamics is of a spin-glass type and cannot be
described by
a finite set of closed dynamic equations.}.  In subsequent
calculations we will, however, assume the number of arguments
$(x,y)$
for which $P[x,y;\bJ]$ is to be evaluated (and thus our
number of
order parameters) to go to infinity only after the limit
$N\to\infty$
has been taken. This will eliminate many technical subtleties
and will
allow us to use the Fokker-Planck equation
(\ref{eq:fokkerplanck}).

The observables (\ref{eq:learningobservables}) are indeed
of the
mean-field type in the sense of (\ref{eq:meanfield}).
Insertion into
(\ref{eq:meanfield}) immediately shows this to be true
for the scalar
observables $Q[\bJ]$ and $R[\bJ]$.  Checking (\ref{eq:meanfield})
for
the function $P[x,y;\bJ]$ is less trivial. Here we have to
use the
property that $\set$ has been composed in a random manner.
We denote
with $\cal{I}$ the set of all {\em different} indices in the
list
$(i_1,\ldots,i_\ell)$, with $n_k$ giving the number of times a
number
$k$ occurs, and with ${\cal I}^\pm\subseteq{\cal I}$ defined as
the
set of all indices $k\in{\cal I}$ for which $n_k$ is even ($+$),
or
odd ($-$). Note that with these definitions
$\ell=\sum_{k\in{\cal I}^+} n_k+
\sum_{k\in{\cal I}^{-}}n_k\geq 2|{\cal
I}^+| +|{\cal I}^-|$.  We then obtain the following scaling
identity:
\begin{eqnarray}
\bra \xi_{i_1}\ldots\!\! & \!\!\xi_{i_\ell}\!\!& \!\!\!
\delta[x\minus\bJ\cdot\bxi] \ \delta[y\minus\bB\cdot\bxi]
\ket_{\ketset}
\nonumber
\\
%~~~~~~~~~~~~~~~~~~~~~~~~~~~~~~~~~~~~~~~~~~~~~~~
%\ed
%\bd
%~~~~~~~~~~~~~~~~~
&=&\! i^\ell\int\frac{d\hat{x} \ d\hat{y}}{(2\pi)^2}
e^{i[x\hat{x}+y\hat{y}]}
\bigbra
\left[\prod_{k\in {\cal I}}\xi_k^{n_k}e^{-i\xi_k
[\hat{x}J_k+\hat{y}B_k]}\right]
\left[\prod_{k\notin {\cal I}}e^{-i\xi_k
[\hat{x}J_k+\hat{y}B_k]}\right]
\bigket_{\!\!\ketset}  \nonumber \\
%\ed
%\be
&=& \! i^\ell\int\frac{d\hat{x} \ d\hat{y}}{(2\pi)^2}
e^{i[x\hat{x}+y\hat{y}]}
\order\left(N^{-\frac{1}{2}|{\cal I}^-|}\right)
=
\order\left(N^{-\frac{1}{2}|{\cal I}^-|}\right)
\label{eq:nastyscaling}
\end{eqnarray}
This gives for derivatives of $P[x,y;\bJ]$:
\bd
\frac{\partial^\ell}{\partial J_{i_1}\!\ldots\!\partial
J_{i_\ell}}
P[x,y;\bJ]=(\minus 1)^\ell\frac{\partial^\ell}{\partial
x^\ell}
\bra \xi_{i_1}\!\ldots\! \xi_{i_\ell} \
\delta[x\minus\bJ\cdot\bxi] \ \delta[y\minus\bB\cdot\bxi]
\ket_{\ketset}
=
\order\left(N^{-\frac{1}{2}|{\cal I}^-|} \right)
\ed
Since $\frac{1}{2}\ell \minus |{\cal I}|\plus
\frac{1}{2}|{\cal I}^-|=
\frac{1}{2}[\ell\minus |{\cal I}^-| \minus 2|{\cal I}^+|]
\geq 0$, the
function $P[x,y;\bJ]$ is indeed found to be a mean-field
 observable.

\subsection{Deterministic Dynamical Laws}

We next show that for the observables
(\ref{eq:learningobservables})
the diffusion matrix elements $G_{\mu\nu}^{\star\star\star}$
in the
Fokker-Planck equation (\ref{eq:fokkerplanck}) vanish for
$N\to\infty$. Our observables will consequently obey
deterministic
dynamical laws.  The diffusion terms associated with
$Q[\bJ]$ and
$R[\bJ]$ are trivial. For on-line learning we find:
\begin{eqnarray*}
\left[\!\!\begin{array}{ll}
G^{\rm onl}_{QQ}[\ldots]\\
G^{\rm onl}_{QR}[\ldots]\\
G^{\rm onl}_{RR}[\ldots]
\end{array}\!\!\right]
&=&
\frac{\eta^2}{N} \bigbra\!\!\bigbra \sum_{ij}
\xi_i\xi_j \ \G^2[\bJ\inn\bxi,\bB\inn\bxi]
\left[\!\!\begin{array}{c}
(2J_i)(2J_j)\\
(2J_i)(B_j)\\
(B_i)(B_j)
\end{array}\!\!\right]
\bigket_{\!\!\ketset}\bigket_{\!\!\ketop;t}
\\
&=&\frac{\eta^2}{N}\int\!dx \ dy~P[x,y] \ \G^2[x,y]
\left[\!\!\begin{array}{c}
4 x^2
\\
2xy
\\
y^2
\end{array}\!\!\right]=\order\left(\frac{1}{N}\right) \ ,
\end{eqnarray*}
where the notation $\bra\cdots \ket_{\ketop;t}$ refers to
sub-shell averages
with respect to the order parameters $Q$, $R$ and
$\{P[x,y]\}$ at time
$t$. For batch learning, similarly:
\bd \left[\!\!\begin{array}{ll} G^{\rm bat}_{QQ}[\ldots]\\
G^{\rm
bat}_{QR}[\ldots]\\ G^{\rm bat}_{RR}[\ldots]
\end{array}\!\!\right]
=
\frac{\eta^2}{N} \bigbra\!\sum_{ij}
\bra\xi_i\G[\bJ\inn\bxi,\bB\inn\bxi]
\ket_{\ketset}
\bra \
\xi_j \ \G[\bJ\inn\bxi,\bB\inn\bxi]
\ket_{\ketset}
\left[\!\!\begin{array}{c}
(2J_i)(2J_j)\\
(2J_i)(B_j)\\
(B_i)(B_j)
\end{array}\!\!\right]\!
\bigket_{\!\!\ketop;t}
\ed
\bd
=\frac{\eta^2}{N}
\left[\!\!\begin{array}{c}
4\left\{\int\!dx \ dy~P[x,y] \ x \ \G[x,y]
\right\}^2
\\
2
\left\{\room\int\!dx \ dy~P[x;y] \ x \ \G[x,y]
\right\}
\left\{\room\int\!dx \ dy~P[x;y] \ y \ \G[x,y]
\right\}
\\
\left\{\room\int\!dx \ dy~P[x;y] \ y \ \G[x,y]
\right\}^2
\end{array}\!\!\right]=\order\left(\frac{1}{N}\right)
\ed
In calculating diffusion terms which involve the order
parameter
function $P[x;y;\bJ]$ we will again need the scaling
 property
(\ref{eq:nastyscaling}).  First we turn to on-line learning
diffusion
terms with just one occurrence of $P[x,y;\bJ]$
\bd
\left[\!\!\begin{array}{ll} G^{\rm onl}_{Q,P[x,y]}[\ldots]\\
G^{\rm
onl}_{R,P[x,y]}[\ldots]
\end{array}\!\!\right]
=
~~~~~~~~~~~~~~~~~~~~~~~~~~~~~~~~~~~~~~~~~~~~~~~~~~~~~~~~
~~~~~~~~~~~
\ed
\bd
-
\frac{\eta^2}{N} \bigbra\!\!\!\bigbra \sum_{ij}
\xi_i\xi_j \ \G^2[\bJ\inn\bxi,\bB\inn\bxi]
\frac{\partial}{\partial x}
\left[\!\begin{array}{c}
2J_i \ \bra\xi^\prime_j \ \delta[x\minus\bJ\inn\bxi^\prime] \
\delta[y\minus\bB\inn\bxi^\prime]\ket_{\ketset}
\\
B_i \ \bra\xi^\prime_j \ \delta[x\minus\bJ\inn\bxi^\prime] \
\delta[y\minus\bB\inn\bxi^\prime]\ket_{\ketset}
\end{array}\!\right]
\bigket_{\!\!\ketset}\bigket_{\!\!\ketop;t}
\ed
\bd
=
-\eta^2\int\!dx \
dy~P[x,y] \ \G^2[x,y]
\left[\!\!\begin{array}{c}
2x  \\ y \end{array}\!\!\right]
\order\left(\frac{1}{\sqrt{N}}\right)=\order
\left(\frac{1}{\sqrt{N}}\right)
\ed
For batch learning we find a similar result:
\begin{eqnarray*}
\left[\!\begin{array}{ll}
G^{\rm bat}_{Q,P[x,y]}[\ldots]\\
G^{\rm bat}_{R,P[x,y]}[\ldots]
\end{array}\!\right]
&=& -
\frac{\eta^2}{N} \bigbra \sum_{ij}
\bra \xi_i \ \G[\bJ\inn\bxi,\bB\inn\bxi]
\ket_{\ketset}
\bra
\xi_j \ \G[\bJ\inn\bxi,\bB\inn\bxi]
\ket_{\ketset}
%~~~~~~~~~~
\right. \\
%\ed
%\bd
%\left.
%~~~~~~~~~~~~~~~~~~~~~~~~~~
&&~~~~~~\times \left.
\frac{\partial}{\partial x}
\left[\!\begin{array}{c}
(2J_i)
\bra \xi^\prime_j \ \delta[x\minus\bJ\inn\bxi^\prime] \
\delta[y\minus\bB\inn\bxi^\prime]\ket_{\ketset}
\\
(B_i)
\bra \xi^\prime_j \ \delta[x\minus\bJ\inn\bxi^\prime] \
\delta[y\minus\bB\inn\bxi^\prime]\ket_{\ketset}
\end{array}\!\right]
\!\bigket_{\!\!\ketop;t} \\
%\ed
%\bd
&=&
-\eta^2\int\!dx \ dy~P[x,y] \ \G[x,y]
\left[\!\!\begin{array}{c}
2x \\
y\end{array}\!\!\right] \order\left(\frac{1}{\sqrt{N}}
\right)=
\order\left(\frac{1}{\sqrt{N}}\right)
\end{eqnarray*}
The non-trivial terms are those where two derivatives of
 the order
parameter function $P[x,y;\bJ]$ come into play. These are
dealt with
by separating $i=j$ from $i\neq j$ terms, in combination
with
(\ref{eq:nastyscaling}):
\begin{eqnarray*}
G^{\rm onl}_{P[x,y],P[x^\prime,y^\prime]}\!&[\ldots]&\!
=
\frac{\eta^2}{N} \bigbra\bra \sum_{ij}
\xi_i\xi_j \ \G^2[\bJ\inn\bxi,\bB\inn\bxi]\ket_{\ketset}
\frac{\partial^2}{\partial x\partial x^\prime}
\right. \\
%~~~~~~~~~~~~~~~~~~~~
%\ed
%\bd \left.
%~~~~~~~~~~~~~~~~~~~~
&&\!\!\!\!\!\!\!\!\!\!\!\!
\bra\xi_i^\prime \ \delta[x\minus\bJ\inn\bxi^\prime] \
\delta[y\minus\bB\inn\bxi^\prime]\ket_{\ketset} \left.
\bra\xi_j^\prime \ \delta[x^\prime\minus
\bJ\inn\bxi^\prime] \
\delta[y^\prime\minus\bB\inn\bxi^\prime]\ket_{\ketset}
\!\bigket_{\!\!\ketop;t} \\
%\ed
%\bd
&=&\!\!
\frac{\eta^2}{N}\frac{\partial^2}{\partial x
\partial x^\prime}
 \bigbra
\sum_i\order(N^{-1})+\sum_{i\neq j}\order(N^{-2})
\!\bigket_{\!\!\ketop;t}=\order(N^{-1})
\end{eqnarray*}
Similarly:
\begin{eqnarray*}
%\bd
G^{\rm bat}_{P[x,y],P[x^\prime,y^\prime]}\!&[\ldots]&=\!
\frac{\eta^2}{N}\bigbra \sum_{ij}
\bra\xi_i \ \G[\bJ\inn\bxi,\bB\inn\bxi]
\ket_{\ketset}
\bra\xi_j \ \G[\bJ\inn\bxi,\bB\inn\bxi]
\ket_{\ketset} \
\frac{\partial^2}{\partial x\partial x^\prime}
\right. \\
%\ed
%\bd
&&\!\!\!\!\!\!\!\!\!\!\!\!
\bra\xi_i^\prime \ \delta[x\minus\bJ\inn\bxi^\prime] \
\delta[y\minus\bB\inn\bxi^\prime]\ket_{\ketset} \left.
\bra\xi_j^\prime \ \delta[x^\prime\minus
\bJ\inn\bxi^\prime] \
\delta[y^\prime\minus\bB\inn\bxi^\prime]\ket_{\ketset}
\!\bigket_{\!\!\ketop;t} \\
&=&\!\!
\frac{\eta^2}{N}\frac{\partial^2}{\partial x\partial x^\prime}
\bigbra
\sum_{i}\order(N^{-2})+\sum_{i\neq j}\order(N^{-2})
\!\bigket_{\!\!\ketop;t}=\order(N^{-1})
\end{eqnarray*}
All diffusion terms vanish in the limit $N\to\infty$.  The
Fokker-Planck equation (\ref{eq:fokkerplanck}) reduces to
the
Liouville equation
$\frac{d}{dt}P_t(\bOmega)=-\sum_\mu
\frac{\partial}{\partial
\Omega_\mu}[F_\mu[\bOmega;t]P_t(\bOmega)]$, describing
deterministic
evolution for our macroscopic observables:
$\frac{d}{dt}\bOmega=\bF[\bOmega;t]$.  These deterministic
equations
we will now work out explicitly.

\subsubsection*{On-Line Learning}
\vspace*{-2mm}

First we deal with the scalar observables
$Q$ and $R$:
%\bd
\begin{eqnarray}
\frac{d}{dt}Q &=& \lim_{N\to\infty}\left\{
2\eta \ \bra\bra(\bJ\cdot\bxi) \
\G[\bJ\inn\bxi,\bB\inn\bxi]\ket_{\ketset}\ket_{\ketop;t}
+ \eta^2\bra\bra\G^2[\bJ\inn\bxi,\bB\inn\bxi]
\ket_{\ketset}\ket_{\ketop;t}\right\}
\nonumber\\
&=&
2\eta \int\!dx dy~P[x,y] \ x \ \G[x,y]
+ \eta^2\int\!dx dy~P[x,y] \ \G^2[x,y]
\label{eq:onlinedQdt}
\end{eqnarray}
%\ee
\be
\frac{d}{dt}R =
\lim_{N\to \infty}
\eta \ \bra\bra(\bB\inn\bxi) \ \G[\bJ\inn\bxi,\bB\inn\bxi]
\ket_{\ketset}\ket_{\ketop;t}
=
\eta  \int\!dx dy~P[x,y] \ y \ \G[x,y]~~~
\label{eq:onlinedRdt}
\ee
The equations (\ref{eq:onlinedQdt},\ref{eq:onlinedRdt})
are identical
to those found in the $\alpha \rightarrow \infty$
formalism.  The
difference is in the function to be substituted for
$P[x,y]$, which
here is the solution of
\bd
\frac{\partial }{\partial t}P[x,y] = \lim_{N\to\infty}
\left\{
\minus \eta\frac{\partial}{\partial x}
\sum_i \bra\bra\xi_i \ \G[\bJ\inn\bxi,\bB\inn\bxi]
\ket_{\ketset}
\bra \xi^\prime_i  \ \delta[x\minus \bJ\inn\bxi^\prime]
\delta[y\minus \bB\inn\bxi^\prime]\ket_{\ketset}
\ket_{\ketop;t}
\right.
\ed
\bd
\left.
+ \frac{\eta^2}{2N}\frac{\partial^2}{\partial x^2}
\bra\sum_{ij}\bra
\xi_i\xi_j \ \G^2[\bJ\inn\bxi,\bB\inn\bxi]\ket_{\ketset}
\bra \xi^\prime_i\xi_j^\prime \
\delta[x\minus \bJ\inn\bxi^\prime] \
\delta[y\minus \bB\inn\bxi^\prime]\ket_{\ketset}
\ket_{\ketop;t}
\room\right\}
\ed
According to (\ref{eq:nastyscaling}) the off-diagonal
terms $i\neq j$
in the contribution with the second derivative
$\frac{\partial^2}{\partial x^2}$ together contribute
only vanishing
orders ($N^{-1}$), so that we need only consider the
diagonal $i=j$
ones:
\bd
\frac{\partial }{\partial t}P[x,y] = \lim_{N\to\infty}
\left\{
\minus \eta\frac{\partial}{\partial x} \bra\bra\bra
\G[\bJ\inn\bxi,\bB\inn\bxi] \ (\bxi\inn\bxi^\prime) \
\delta[x\minus
\bJ\inn\bxi^\prime] \ \delta[y\minus
\bB\inn\bxi^\prime]\ket_{\ketset}\ket_{\ketset}
\ket_{\ketop;t} \right.
\vspace*{-2mm}
\ed
\bd
\left.  ~~~~~~~~~~~~~~~~~~~~ +
\frac{1}{2}\eta^2\frac{\partial^2}{\partial x^2} \bra
P[x,y;\bJ] \
\bra\G^2[\bJ\inn\bxi,\bB\inn\bxi]\ket_{\ketset}
\ket_{\ketop;t} \room\right\}
\ed
\be
= -\eta\frac{\partial}{\partial x}\int\!dx^\prime
dy^\prime \
\G[x^\prime,y^\prime]  \C[x,y;x^\prime,y^\prime]
+ \frac{1}{2}\eta^2\!\!
\int\!dx^\prime dy^\prime \ P[x^\prime,y^\prime]
\G^2[x^\prime,y^\prime]
~ \frac{\partial^2}{\partial x^2}P[x,y]
\label{eq:onlinedPdt}
\ee
with the function
\be
\C[x,y;x^\prime,y^\prime]=
\lim_{N\to\infty}
 \bra\bra\bra
\delta[x\minus \bJ\cdot\bxi] \ \delta[y\minus \bB\cdot\bxi] \
(\bxi\cdot\bxi^\prime) \
\delta[x^\prime\minus\bJ\cdot\bxi^\prime] \
\delta[y^\prime\minus\bB\cdot\bxi^\prime]
\ket_{\ketset}\ket_{\ketset}
\ket_{\ketop;t}
\label{eq:green}
\ee

\subsubsection*{Batch Learning}
\vspace*{-2mm}

Here we can again use the scaling relation
(\ref{eq:nastyscaling})
to eliminate terms. For $Q$ and $R$ one finds
\bd
\frac{d}{dt}Q=\lim_{N\to \infty}\left\{
2\eta \ \bra\bra(\bJ\inn\bxi) \ \G[\bJ\inn\bxi,
\bB\inn\bxi]\ket_{\ketset}
\ket_{\ketop;t}
+ \frac{\eta^2}{N}\bra\sum_{i}
\bra\xi_i \ \G[\bJ\!\inn\!\bxi,\bB\!\inn\!\bxi]
\ket_{\ketset}^2
\ket_{\ketop;t}
\right\}
\ed
\be
=
2\eta \int\!dx dy~P[x,y] \ x \ \G[x;y]
\label{eq:batchdQdt}
\ee
\bd
\frac{d}{dt}R=\lim_{N\to\infty}
\eta \ \bra
\bra(\bB\inn\bxi) \ \G[\bJ\inn\bxi,\bB\inn\bxi]
\ket_{\ketset}
\ket_{\ketop;t}
~~~~~~~~~~~~~~~~~~~~~~~~~~~~~~~~~~~~~~~~~~
\ed
\be
=\eta\int\!dx dy~ P[x,y] \ y \ \G[x;y]
\label{eq:batchdRdt}
\ee
Finally we calculate the temporal derivative of the
joint field
distribution:
\bd
\frac{\partial }{\partial t}P[x,y] = \lim_{N\to\infty}
\left\{\room
\minus \eta \frac{\partial}{\partial x}
\bra\bra\bra  \G[\bJ\inn\bxi,\bB\inn\bxi]
(\bxi\inn\bxi^\prime)  \delta[x\minus\bJ\inn\bxi^\prime]
\delta[y\minus\bB\inn\bxi^\prime]
\ket_{\ketset}\ket_{\ketset}
\ket_{\ketop;t}
\right.
\ed
\bd
\left.
+\frac{\eta^2}{2N}\frac{\partial^2}{\partial x^2}
\bra\sum_{ij}
\bra\xi_i  \G[\bJ\!\inn\!\bxi,\bB\!\inn\!\bxi]
\ket_{\ketset}
\bra\xi_j  \G[\bJ\!\inn\!\bxi,\bB\!\inn\!\bxi]
\ket_{\ketset}
\bra \xi_i^\prime \xi_j^\prime
\delta[x\minus\bJ\!\inn\!\bxi^\prime]
\delta[y\minus\bB\!\inn\!\bxi^\prime]
\ket_{\ketset}
\ket_{\ketop;t}\room
\right\}
\ed
\be
=
-\eta \frac{\partial}{\partial x}\int\!dx^\prime dy^\prime~
\G[x^\prime,y^\prime] \ \C[x,y;x^\prime,y^\prime]
\label{eq:batchdPdt}
\ee
\normalsize
The difference between the macroscopic equations for
batch and on-line
learning is merely the presence (on-line) or absence
(batch) of those
terms which are quadratic in the learning rate $\eta$.

\subsubsection*{Resume}
\vspace*{-2mm}

We separate the function $\cC[x,y;x^\prime,y^\prime]$
(\ref{eq:green}),
which plays a role similar to that of a Green's function,
into two terms ($\bxi=\bxi^\prime$ versus
$\bxi\neq\bxi^\prime$):
\bea
&\hspace{-1em}&\cC[x,y;x^\prime,y^\prime]=
\alpha^{-1}\delta[x\minus x^\prime] \
\delta[y\minus y^\prime] \ P[x,y]
+\cA[x,y;x^\prime,y^\prime]
\nonumber\room
\\
%\ed
%\bd
&\hspace{-1em}&\cA[x,y;x^\prime,y^\prime]=
\lim_{N\to\infty}\room
\label{eq:greensfunction}\\
%~~~~~~~~~~~~~~~~~~~~~~~~~~~~~~~~~~~~~~~~~~~~~~~~~~~~~~~~~~~~~~~~~~~~~~~~~~~~~~~~~~~~~~~~~~~~~~~~
%\ed
%\be
&\hspace{-1em}&\bra\bra\bra [1\minus
\delta_{\bxi \bxi^\prime}]
\delta[x\minus \bJ\inn\bxi] \ \delta[y\minus \bB\inn\bxi] \
(\bxi\inn\bxi^\prime) \
\delta[x^\prime\minus\bJ\inn\bxi^\prime] \
\delta[y^\prime\minus\bB\inn\bxi^\prime]
\ket_{\ketset}\ket_{\ketset}
\ket_{\ketop;t} \nonumber
\eea
With this definition our macroscopic laws, which are
exact for
$N\to\infty$ but not yet closed due to the appearance
of the
microscopic probability density $p_t(\bJ)$ in the sub-shell
average of
(\ref{eq:greensfunction}), can be summarised as follows
\bea
&\hspace{-1em}&\frac{d}{dt}Q= 2\eta \int\!dx  dy~P[x,y]
\ x \ \G[x;y] +
\kappa\eta^2\int\!dx  dy~ P[x,y] \ \G^2[x;y]
\label{eq:dQdtfurther} \\
%\ee
%\be
&\hspace{-1em}&\frac{d}{dt}R= \eta  \int\!dx  dy~ P[x,y]
\ y \ \G[x;y]
\label{eq:dRdtfurther} \\
%\ee
%\bd
&\hspace{-1em}&\frac{\partial }{\partial t}P[x,y]=
-\frac{\eta}{\alpha}\frac{\partial}{\partial x}\left[ \G[x,y]
\ P[x,y]\right]
-\eta\frac{\partial}{\partial x}\int\!dx^\prime dy^\prime \
\G[x^\prime,y^\prime]  \ \cA[x,y;x^\prime,y^\prime]
\nonumber \\
%\ed
%\be
&& \hspace{1cm} + \frac{1}{2}\kappa\eta^2
\int\!dx^\prime dy^\prime \ P[x^\prime,y^\prime] \
\G^2[x^\prime,y^\prime]
~\frac{\partial^2}{\partial x^2}P[x,y]
\label{eq:dPdtfurther}
\eea
in which $\kappa=1$ for on-line learning and $\kappa=0$
for batch
learning. The complexity of the problem is fully concentrated
in
the Green's function $\cA[x,y;x^\prime,y^\prime]$.

\subsection{Closure of Macroscopic Dynamical Laws}

We now close our macroscopic laws by making, for $N\to\infty$,
the two
key assumptions underlying dynamical replica theories:
\begin{enumerate}
\item
Our macroscopic observables $\{Q,R,P\}$  obey {\em closed}
dynamic equations.
\item
These macroscopic equations are self-averaging with respect
to the
disorder, i.e. the microscopic realisation of the training
set $\set$.
\end{enumerate}
Assumption 1 implies that all microscopic probability
variations
within the $\{Q,R,P\}$ subshells of the $\bJ$-ensemble
are either
absent or irrelevant to the evolution of $\{Q,R,P\}$.
We may
consequently make the simplest self-consistent choice
for $p_t(\bJ)$
in evaluating the macroscopic laws, i.e. in
(\ref{eq:greensfunction}):
microscopic probability equipartitioning in
the $\{Q,R,P\}$-subshells
of the ensemble, or
\be
p_t(\bJ)~\to ~w(\bJ)\sim
\delta[Q\minus Q[\bJ]] \
\delta[R\minus R[\bJ]] \ \prod_{xy}\delta[P[x,y]
\minus P[x,y;\bJ]]
\label{eq:equipart}
\ee
This distribution depends on time via the order
parameters
$\{Q,R,P\}$.
Note that (\ref{eq:equipart})
leads to exact macroscopic laws if our observables
$\{Q,R,P\}$
for $N\to\infty$ indeed obey closed equations,
and is true
in equilibrium for detailed balance models in
which the  Hamiltonian
can be written in terms of $\{Q,R,P\}$.
It is an approximation if our observables do not
obey closed
equations. Assumption 2 allows us to average the macroscopic
laws over the
disorder; for mean-field models it is
usually convincingly supported  by numerical
simulations, and
can be proven using the path integral formalism
(see e.g. Horner,
1992). We write averages over all training
sets $\set\subseteq \{-1,1\}^N$, with $|\set|=p$, as $\bra
\ldots\ket_{ \ \rmsets}$.
Our assumptions result in the closure of
(\ref{eq:dQdtfurther},\ref{eq:dRdtfurther},\ref{eq:dPdtfurther}),
since now the function $\A[x,y;x^\prime,y^\prime]$ is expressed
fully in
terms of $\{Q,R,P\}$:
\bd
\A[x,y;x^\prime,y^\prime]=
\lim_{N\to\infty}
~~~~~~~~~~~~~~~~~~~~~~~~~~~~~~~~~~~~~~~~~~~~~~~~~~~~~~~~~~~~~~~~
\ed
%%%%%%%%%%%% The first ket was too tight
\bd
\bigbra\!
\frac{\int\!d\bJ~w(\bJ) \
\bra\!\bra\delta[x\minus\bJ\inn\bxi] \ \delta[y\minus\bB\inn\bxi]
(\bxi\inn\bxi^\prime)  [1\minus\delta_{\bxi\bxi^\prime}] \
\delta[x^\prime\minus\bJ\inn\bxi^\prime] \
\delta[y^\prime\minus\bB\inn\bxi^\prime]
\ket_{\ketset}\ket_{\ketset}}
{\int\!d\bJ~w(\bJ)}\!\bigket_{\!\rmsets}
\ed
The final ingredient of dynamical replica theory is the
realization
that averages of fractions can be calculated with the replica
identity
\bd
\bigbra \frac{\int\!d\bJ~W[\bJ,\bz]G[\bJ,\bz]}
{\int\!d\bJ~W[\bJ,\bz]}\bigket_{\!\!\bz}=
\lim_{n\to 0} \int\!d\bJ^1\cdots d\bJ^n~\bra G[\bJ^1,\bz]
\prod_{\alpha=1}^n W[\bJ^\alpha,\bz]\ket_{\bz}
\ed
giving
\bd
\A[x,y;x^\prime,y^\prime]=\lim_{N\to\infty}\lim_{n\to 0}
\int\!\prod_{\alpha=1}^n w(\bJ^\alpha) \ d\bJ^\alpha
~~~~~~~~~~~~~~~~~~~~~~~~~~~~~~~~~~~~~~~~~\vspace*{-1mm}
\ed
\bd
\bigbra
\bra\bra\delta[x\minus\bJ^1\inn\bxi] \ \delta[y\minus
\bB\inn\bxi] \
(\bxi\inn\bxi^\prime) \ [1\minus\delta_{\bxi\bxi^\prime}] \
\delta[x^\prime\minus\bJ^1\inn\bxi^\prime] \
\delta[y^\prime\minus\bB\inn\bxi^\prime]
\ket_{\ketset}\ket_{\ketset}
\bigket_{\rmsets}
\ed
Since each weight component scales as $J^\alpha_i=
\order(N^{-\frac{1}{2}})$ we transform variables in such a
way that
our calculations will involve $\order(1)$ objects:
\bd
(\forall i)(\forall\alpha):~~~~~~
J^\alpha_i=(Q/N)^{\frac{1}{2}}\sigma^\alpha_i,~~~~
B_i=N^{-\frac{1}{2}}\tau_i
\ed
This ensures
$\sigma_i^\alpha=\order(1)$, $\tau_i=\order(1)$, and
reduces various
constraints to ordinary spherical ones: $(\bsigma^\alpha)^2
=\btau^2=N$
for all $\alpha$.  Overall prefactors generated by these
transformations always vanish due to $n\to 0$. We find a
new effective
measure: $\prod_{\alpha=1}^n w(\bJ^\alpha) \ d\bJ^\alpha
\rightarrow
\prod_{\alpha=1}^n \tilde{w}(\bsigma^\alpha) \ d\bsigma^\alpha$,
with \be
\tilde{w}(\bsigma) \sim \delta\left[N\minus \bsigma^2\right]
\delta\left[NRQ^{-\frac{1}{2}}\minus\btau\inn \bsigma\right]
\prod_{xy}\delta\left[P[x,y]\minus
P[x,y;(Q/N)^{\frac{1}{2}}\bsigma]\right]
\label{eq:newmeasure}
\ee
We thus arrive at
\bd
\A[x,y;x^\prime,y^\prime]=\lim_{n\to 0}\lim_{N\to\infty}
\int\!\prod_{\alpha=1}^n \tilde{w}(\bsigma^\alpha) \
d\bsigma^\alpha
\bigbra\room
\bigbras
(\bxi^\prime\cdot\bxi)[1\minus\delta_{\bxi\bxi^\prime}]
~~~~~~~~~~~~~~~~~~~~~~~~~~~
\right.\right.\right.
\ed
\be
\left.\left.\left.
\delta\left[x\minus
\frac{\sqrt{Q}\bsigma^1\inn\bxi}{\sqrt{N}}\right]
\delta\left[y\minus \frac{\btau\inn\bxi}{\sqrt{N}}\right]
\delta\left[x^\prime\minus
\frac{\sqrt{Q}\bsigma^1\inn\bxi^\prime}{\sqrt{N}}
\right]
\delta\left[y\minus
\frac{\btau\inn\bxi^\prime}{\sqrt{N}}\right]
\bigket_{\ketset}\bigket_{\ketset}
\bigket_{\rmsets}
\label{eq:Ainreplicas}
\ee
In the same fashion one can also express $P[x,y]$ in replica
form
(which will prove useful for normalization purposes and for
self-consistency tests):
\be
P_t[x,y]=\lim_{n\to 0}\lim_{N\to\infty}
\int\!\prod_{\alpha=1}^n \tilde{w}(\bsigma^\alpha)
d\bsigma^\alpha
\bigbras
\delta\left[x\minus \frac{\sqrt{Q}\bsigma^1\inn\bxi}{\sqrt{N}}
\right]
\delta\left[y\minus \frac{\btau\inn\bxi}{\sqrt{N}}\right]
\bigket_{\ketset}
\bigket_{\rmsets}
\label{eq:Pinreplicas}
\ee

%%%%%%%%%%%%%%%%%%%%%%%%%%%%%%%%%%%%%%%%%%%%%%%%%%

\section{Replica Calculation of the Green's Function}

\subsection{Disorder Averaging}

In order to perform the disorder average we insert integral
representations for the $\delta$-functions which define the
fields
$(x,y,x^\prime,y^\prime)$ and for the $\delta$-functions in
the
measure (\ref{eq:newmeasure}) which involve $P[x,y]$,
generating $n$
conjugate order parameter functions $\hat{P}_\alpha(x,y)$.
Upon also
writing averages over the training set in terms of the $p$
constituent
vectors $\{\bxi^\mu\}$ we obtain for (\ref{eq:Ainreplicas})
and
(\ref{eq:Pinreplicas}):
\bd
\A[x,y;x^\prime,y^\prime]
=\int\!\frac{d\hat{x} \ d\hat{x}^\prime d\hat{y} \
d\hat{y}^\prime}{(2\pi)^4}e^{i[x\hat{x}+x^\prime\hat{x}^\prime
+y\hat{y}+y\hat{y}^\prime]}
\lim_{n\to 0}\lim_{N\to\infty}
\int\!\prod_{\alpha=1}^n\prod_{x^\pprime
y^\pprime}d\hat{P}_\alpha(x^\pprime,y^\pprime)
~~~~~~~~
\ed
\bd
\int\!
\prod_{\alpha=1}^n \left\{d\bsigma^\alpha \
\delta\left[N\minus (\bsigma^\alpha)^2\right]
\delta\left[\frac{NR}{\sqrt{Q}}\minus\btau\inn
\bsigma^\alpha\right]
e^{iN\int dx^\pprime
dy^\pprime \ \hat{P}_\alpha(x^\pprime,y^\pprime) \
P_t(x^\pprime,y^\pprime)}
\right\}
\ed
\be
\bigbra \frac{1}{p^2}\!
\sum_{\mu\neq \nu}
(\bxi^\mu\!\!\inn\!\bxi^\nu)
e^{\!\!-\frac{i}{\alpha}\sum_{\alpha\lambda}
\hat{P}_\alpha\left(\!\frac{\sqrt{Q}
\bsigma^\alpha\cdot\bxi^\lambda}{\sqrt{N}},
\frac{\btau\cdot\bxi^\lambda}{\sqrt{N}}\!\right)
-\frac{i}{\sqrt{N}}\bxi^\mu\!\cdot[\hat{x}
\sqrt{Q}\bsigma^1+\hat{y}\btau]
-\frac{i}{\sqrt{N}}\bxi^\nu\!\cdot[\hat{x}^\prime
\sqrt{Q}\bsigma^1+\hat{y}^\prime\btau]}
\bigket_{\!\rmsets}
\label{eq:intermediate1}
\ee
\bd
P[x,y]
=\int\!\frac{d\hat{x} \ d\hat{y}}{(2\pi)^2}
e^{i[x\hat{x}+y\hat{y}]}
\lim_{n\to 0}\lim_{N\to\infty}
\int\!\prod_{\alpha=1}^n
\prod_{x^\pprime y^\pprime}
d\hat{P}_\alpha(x^\pprime,y^\pprime)
~~~~~~~~~~~~~~~~~~~~~~~~~~~~~~~~~~~~~~~~~~~
\ed
\bd
\int\!
\prod_{\alpha=1}^n \left\{d\bsigma^\alpha
\delta\left[N\minus (\bsigma^\alpha)^2\right]
\delta\left[\frac{NR}{\sqrt{Q}}\minus\btau\inn
\bsigma^\alpha\right]
e^{iN\int dx^\pprime
dy^\pprime \ \hat{P}_\alpha(x^\pprime,y^\pprime)
\ P_t(x^\pprime,y^\pprime)}
\right\}
\ed
\be
\bigbra \frac{1}{p}
\sum_{\mu=1}^p
e^{-\frac{i}{\alpha}\sum_{\alpha\lambda}
\hat{P}_\alpha\left(\frac{\sqrt{Q}\bsigma^\alpha
\cdot\bxi^\lambda}{\sqrt{N}},
\frac{\btau\cdot\bxi^\lambda}{\sqrt{N}}\right)
-\frac{i}{\sqrt{N}}\bxi^\mu\!\cdot[\hat{x}\sqrt{Q}
\bsigma^1+\hat{y}\btau]}
\bigket_{\rmsets}
\label{eq:intermediate2}
\ee
In calculating the averages over the training sets $\bra
\ldots\ket_{\ \rmsets}$ that occur in (\ref{eq:intermediate1})
and
(\ref{eq:intermediate2}) one can use permutation symmetries
with
respect to sites and pattern labels, leading to the following
compact
results:
\bd
\bigbra \frac{1}{p^2}\!
\sum_{\mu\neq \nu}
(\bxi^\mu\!\!\inn\!\bxi^\nu)
e^{\!\!-\frac{i}{\alpha}\sum_\alpha\sum_{\lambda}
\hat{P}_\alpha\left(\!\frac{\sqrt{Q}\bsigma^\alpha
\cdot\bxi^\lambda}{\sqrt{N}},
\frac{\btau\cdot\bxi^\lambda}{\sqrt{N}}\!\right)
-\frac{i}{\sqrt{N}}\bxi^\mu\!\cdot[\hat{x}\sqrt{Q}
\bsigma^1+\hat{y}\btau]
-\frac{i}{\sqrt{N}}\bxi^\nu\!\cdot[\hat{x}^\prime
\sqrt{Q}\bsigma^1+\hat{y}^\prime\btau]}
\bigket_{\!\rmsets}
\vspace*{-2mm}
\ed
\be
=e^{p\log \cD[0,0]}~\frac{1}{N}\sum_j
\frac{\cE_j[\hat{x},\hat{y}]\cE_j[\hat{x}^\prime,
\hat{y}^\prime]}
{\cD^{2}[0,0]}
+\order(N^{-\frac{1}{2}})
\label{eq:Aworkedout}
\ee
and
\bd
\bigbra \frac{1}{p}
\sum_{\mu=1}^p
e^{-\frac{i}{\alpha}\sum_\alpha\sum_{\lambda}
\hat{P}_\alpha\left(\frac{\sqrt{Q}
\bsigma^\alpha\cdot\bxi^\lambda}{\sqrt{N}},
\frac{\btau\cdot\bxi^\lambda}{\sqrt{N}}\right)
-\frac{i}{\sqrt{N}}\bxi^\mu\!\cdot[\hat{x}
\sqrt{Q}\bsigma^1+\hat{y}\btau]}
\bigket_{\rmsets}
\vspace*{-2mm}
\ed
\be
=e^{p\log \cD[0,0]}
\frac{\cD[\hat{x},\hat{y}]}
{\cD[0,0]}
+\order\left(N^{-\frac{1}{2}}\right)
\label{eq:Pworkedout}
\ee
in which
\begin{eqnarray*}
\cD[u,v] \!&=&\!
\bigbra
e^{-\frac{i}{\alpha}\sum_\alpha
\hat{P}_\alpha\left(\frac{\sqrt{Q}\bsigma^\alpha
\cdot\bxi}{\sqrt{N}},
\frac{\btau\cdot\bxi}{\sqrt{N}}\right)
-\frac{i}{\sqrt{N}}\bxi\cdot[u\sqrt{Q}\bsigma^1+v\btau]}
\bigket_{\bxi} \\
%\ed
%\bd
\cE_j[u,v]\! &=&\!
\bigbra \sqrt{N}\xi_j~
e^{-\frac{i}{\alpha}\sum_\alpha
\hat{P}_\alpha\left(\frac{\sqrt{Q}
\bsigma^\alpha\cdot\bxi}{\sqrt{N}},
\frac{\btau\cdot\bxi}{\sqrt{N}}\right)
-\frac{i}{\sqrt{N}}\bxi\cdot[u\sqrt{Q}\bsigma^1
+v\btau]}
\bigket_{\bxi}
\end{eqnarray*}
and with the abbreviation $\bra f[\bxi]\ket_{\bxi}=
2^{-N}\sum_{\bxi\in\{-1,1\}^N}f[\bxi]$.  These quantities
(which are
both $\order(1)$ for $N\to\infty$) are, in turn, evaluated
by using
the central limit theorem, which ensures that for $N\to\infty$
the $n$
rescaled inner products $\bsigma^\alpha\cdot\bxi/\sqrt{N}$
and the
rescaled inner product $\btau\cdot\bxi/\sqrt{N}$ will
become
(correlated) zero-average Gaussian variables.  After some
algebra one
finds
\bd \cL[u,v;u^\prime,v^\prime]=\frac{1}{N}\sum_j
\cE_j[u,v] \ \cE_j[u^\prime,v^\prime]=
~~~~~~~~~~~~~~~~~~~~~~~~~~~~~~~~~~~~~~~~~
~~~~~~~~~~~~~~~~~~~~~~~~~~
\vspace*{-3mm}
\ed
\bd
=
-Q\sum_{\alpha\beta}q_{\alpha\beta}(\{\bsigma\})
\left[\frac{1}{\alpha}\cF^\alpha_1[u,v]+u \
\delta_{\alpha 1}\cD[u,v]\right]
\left[\frac{1}{\alpha}\cF^\beta_1[u^\prime,v^\prime]+
u^\prime \ \delta_{\beta 1}\cD[u^\prime,v^\prime]\right]
\ed
\bd
-R\sum_{\alpha\beta}
\left[\frac{1}{\alpha}\cF^\alpha_1[u,v]+u \
\delta_{\alpha 1}\cD[u,v]\right]
\left[\frac{1}{\alpha}\cF^\beta_2[u^\prime,v^\prime]+
v^\prime \ \delta_{\beta 1}\cD[u^\prime,v^\prime]\right]
\ed
\bd
-R\sum_{\alpha\beta}
\left[\frac{1}{\alpha}\cF^\alpha_1[u^\prime,v^\prime]+
u^\prime \ \delta_{\alpha 1}\cD[u^\prime,v^\prime]\right]
\left[\frac{1}{\alpha}\cF^\beta_2[u,v]+v \ \delta_{\beta 1}
\cD[u,v]\right]
\ed
\be
-\sum_{\alpha\beta}
\left[\frac{1}{\alpha}\cF^\alpha_2[u,v]+v \ \delta_{\alpha 1}
\cD[u,v]\right]
\left[\frac{1}{\alpha}\cF^\beta_2[u^\prime,v^\prime]+v^\prime
\ \delta_{\beta 1}\cD[u^\prime,v^\prime]\right]
+\order(N^{-\frac{1}{2}})
\label{eq:nastyterm}
\ee
in which $\cD[u,v]$ and the $\cF^\alpha_\lambda[u,v]$ are
given by
 $n\plus 1$ dimensional integrals:
\be
\cD[u,v]=
\int\!\frac{d\bx \ dy~
{\rm det}^{\frac{1}{2}}\!\bA}{(2\pi)^{(n+1)/2}}~
e^{-\frac{1}{2}
\left(\!\!\begin{array}{c}\bx\\ y\end{array}\!\!\right)
\inn\bA
\left(\!\!\begin{array}{c}\bx\\ y\end{array}\!\!\right)
-\frac{i}{\alpha}\sum_\alpha
\hat{P}_\alpha(\sqrt{Q}x_\alpha,y)
-i[u\sqrt{Q}x_1 +vy]}~~~~
\label{eq:Dasintegral}
\ee
\bd
\cF_\lambda^\alpha[u,v]=
~~~~~~~~~~~~~~~~~~~~~~~~~~~~~~~~~~~~~~~~~~~~~~~~~~~~~
~~~~~~~~~~~~~~~~~~~~~~~~~~~~~~
\vspace*{-2mm}
\ed
\be
\int\!\frac{d\bx \ dy~
{\rm det}^{\frac{1}{2}}\!\bA}{(2\pi)^{(n+1)/2}}~
\partial_\lambda\hat{P}_\alpha
(\sqrt{Q}x_\alpha,y)
~e^{-\frac{1}{2}
\left(\!\!\begin{array}{c}\bx\\ y\end{array}\!\!\right)\inn\bA
\left(\!\!\begin{array}{c}\bx\\ y\end{array}\!\!\right)
-\frac{i}{\alpha}\sum_\alpha
\hat{P}_\alpha(\sqrt{Q}x_\alpha,y)
-i[u\sqrt{Q}x_1 +vy]}
\label{eq:Fasintegral}
\ee
with $\lambda\in\{1,2\}$.
The matrix $\bA$ in (\ref{eq:Dasintegral},\ref{eq:Fasintegral})
is
defined by
\be
\bA^{-1}=\left(
\!\!\begin{array}{cccc} q_{11} & \!\!\cdots\!\! &
q_{1n} & \!\! R/\sqrt{Q}\\
\vdots && \vdots & \vdots \\
q_{n1} & \!\!\cdots\!\! & q_{nn} & \!\!R/\sqrt{Q} \\
R/\sqrt{Q} & \!\!\cdots\!\! & R/\sqrt{Q} & 1
\end{array}\!\!\right)
~~~~~~~
q_{\alpha\beta}(\{\bsigma\})=
\frac{1}{N}\sum_i\sigma_i^\alpha \sigma_i^\beta
\label{eq:spinglassops}
\ee
Note that the quantities
(\ref{eq:Dasintegral},\ref{eq:Fasintegral})
depend on the microscopic variables $\bsigma^\alpha$ only
through the
spin-glass order parameters $q_{\alpha\beta}(\{\bsigma\})$.

\subsection{Derivation of Saddle-Point Equations}

We combine the results
(\ref{eq:Aworkedout},\ref{eq:Pworkedout},\ref{eq:nastyterm})
with
(\ref{eq:intermediate1},\ref{eq:intermediate2}).
We use integral
representations for the remaining $\delta$-functions, and
isolate the
$q_{\alpha\beta}$, by inserting
\bd
1=\int\frac{d\bq \ d\hbq \ d\hbQ \ d\hbR}{(2\pi/N)^{n^2+2n}}
~e^{iN[\sum_\alpha(\hQ_\alpha+\hR_\alpha
R/\sqrt{Q})+\sum_{\alpha\beta}\hat{q}_{\alpha\beta}
q_{\alpha\beta}]}
~~~~~~~~~~~~~~~~~~~~~~~~
\vspace*{-2mm}
\ed
\bd
~~~~~~~~~~~~~~~~~~~~~~~~~~\times~
e^{-i\sum_i[\sum_\alpha(\hQ_\alpha(\sigma_i^\alpha)^2+
\hR_\alpha\tau_i\sigma_i^\alpha)-
i\sum_{\alpha\beta}\hat{q}_{\alpha\beta}\sigma_i^\alpha
\sigma_i^\beta]}
\ed
We hereby achieve a full factorisation over sites, and
both (\ref{eq:intermediate1}) and (\ref{eq:intermediate2})
can be written in the form of an integral dominated by
saddle-points:
\bd
\A[x,y;x^\prime,y^\prime]
=\int\!\frac{d\hat{x} \ d\hat{x}^\prime d\hat{y} \
d\hat{y}^\prime}{(2\pi)^4}~e^{i[x\hat{x}+
x^\prime\hat{x}^\prime+y\hat{y}+y\hat{y}^\prime]}
~~~~~~~~~~~~~~~~~~~~~~~~~~~~~~~~~~~~~~~~~~~~~~~~~~~~~
\vspace*{-1mm}
\ed
\bd
\lim_{n\to 0}
\lim_{N\to\infty}
\int\! d\bq \ d\hbq \ d\hbQ \ d\hbR \!
\prod_{\alpha x^\pprime
y^\pprime}\! d\hat{P}_\alpha(x^\pprime,y^\pprime)~
e^{N\LPsi[\bq,\hbq,\hbQ,\hbR,\{\hat{P}\}]}
\frac{\cL[\hat{x},\hat{y};\hat{x}^\prime,\hat{y}^\prime]}
{\cD^{2}[0,0]}
\ed
\bd
P[x,y]
=\int\!\frac{d\hat{x} \ d\hat{y}}{(2\pi)^2}~
e^{i[x\hat{x}+y\hat{y}]}
~~~~~~~~~~~~~~~~~~~~~~~~~~~~~~~~~~~~~~~~~~~~
~~~~~~~~~~~~~~~~~~~~~~~
\vspace*{-1mm}
\ed
\bd
\lim_{n\to 0}\lim_{N\to\infty}
\int\! d\bq \ d\hbq \ d\hbQ \ d\hbR
\! \prod_{\alpha x^\pprime
y^\pprime}\! d\hat{P}_\alpha(x^\pprime,y^\pprime)~
e^{N\LPsi[\bq,\hbq,\hbQ,\hbR,\{\hat{P}\}]}
\frac{\cD[\hat{x},\hat{y}]}
{\cD[0,0]}
\ed
with
\bd
\Psi[\ldots]=
i\sum_\alpha(\hQ_\alpha+\hR_\alpha
R/\sqrt{Q})+i\sum_{\alpha\beta}\hat{q}_{\alpha\beta}
\ q_{\alpha\beta}
+i\sum_\alpha\int dx \
dy~\hat{P}_\alpha(x,y)P[x,y]
\vspace*{-1mm}
\ed
\bd
+\alpha \log \cD[0,0]
+\lim_{N\to\infty}\frac{1}{N}\sum_i
\log \int\!d\bsigma
~
e^{-i\sum_\alpha[\hQ_\alpha\sigma_\alpha^2+
\hR_\alpha\tau_i\sigma_\alpha]
-i\sum_{\alpha\beta}\hat{q}_{\alpha\beta}
\sigma_\alpha\sigma_\beta}
\ed
The above expressions for $\cA[x,y;x^\prime,y^\prime]$
and $P[x,y]$ will be given by the intensive parts of the
integrands,
evaluated in the dominating saddle-point of $\Psi$.
We can use the
equation for $P[x,y]$ to verify
that all expressions are properly normalized.
After a simple transformation of
some integration variables,
\bd
\hat{q}_{\alpha\beta}\to\hat{q}_{\alpha\beta}-\hat{Q}_\alpha
\delta_{\alpha\beta}
~~~~~~~~~~~~~
\hat{R}_\alpha \to \sqrt{Q}\hat{R}_\alpha
\ed
we arrive at the simple result
\bea
\A[x,y;x^\prime,y^\prime]\!\!
&=&\!\!\!\int\!\frac{d\hat{x} \ d\hat{x}^\prime
\ d\hat{y} \
d\hat{y}^\prime}{(2\pi)^4}e^{i[x\hat{x}+x^\prime
\hat{x}^\prime+y\hat{y}+y\hat{y}^\prime]}
\lim_{n\to 0}
\frac{\cL[\hat{x},\hat{y};\hat{x}^\prime,\hat{y}^\prime]}
{\cD^{2}[0,0]}
\label{eq:finalA} \\
%\ee
%\be
P[x,y]\!\!&=&\!\!\!\int\!\frac{d\hat{x} \
d\hat{y}}{(2\pi)^2}e^{i[x\hat{x}+y\hat{y}]}
\lim_{n\to 0} \frac{\cD[\hat{x},\hat{y}]}
{\cD[0,0]}
\label{eq:finalP}
\eea
in which all functions are to be evaluated upon choosing
for the order
parameters the appropriate saddle-point of $\Psi$, which
itself takes the
form:
\bd
\Psi[\ldots]=
i\sum_\alpha\hQ_\alpha(1\minus q_{\alpha\alpha})+
iR\sum_\alpha \hR_\alpha
+i\sum_{\alpha\beta}\hat{q}_{\alpha\beta} \ q_{\alpha\beta}
+i\sum_\alpha\int dx
dy~\hat{P}_\alpha(x,y)P[x,y]
\vspace*{-2mm}
\ed
\be
+\alpha \log \cD[0,0]
+\lim_{N\to\infty}\frac{1}{N}\sum_i
\log \int\!d\bsigma
~
e^{-i\tau_i\sqrt{Q}\sum_\alpha \hR_\alpha\sigma_\alpha
-i\sum_{\alpha\beta}\hat{q}_{\alpha\beta}
\sigma_\alpha\sigma_\beta}
\label{eq:finalpsi}
\ee
With $\cD[u,v]$ given by (\ref{eq:Dasintegral}) and with the
function
$\cL[u,v;u^\prime,v^\prime]$ given by (\ref{eq:nastyterm}).
The
auxiliary order parameters $q_{\alpha\beta}$ have the usual
interpretation in terms of the average probability density
for finding
a mutual overlap $q$ of two independently evolving weight
vectors with
the same realization of the training set
(see e.g. M\'{e}zard et al,
1987):
\be
\bra P(q)\ket_{\ \rmsets}=\bigbra  \bigbras
\delta\left[q\minus \frac{\bJ^a\inn\bJ^b}{|\bJ^a||\bJ^b|}
\right]\bigkets  \bigket_{\rmsets}
=~\lim_{n\to 0}\frac{1}{n(n\minus1)}\sum_{\alpha\neq
\beta}\delta[q\minus q_{\alpha\beta}]
\label{eq:interpretation}
\ee

We now make the replica symmetric (RS) ansatz in the
extremisation
problem, which according to (\ref{eq:interpretation})
is equivalent to
assuming ergodicity.  With a modest amount of foresight
we put
\bd
q_{\alpha\beta}=q_0 \delta_{\alpha\beta}+q  [1\minus
\delta_{\alpha\beta}] ~~~~~~~~~~~~
\hat{q}_{\alpha\beta}=\frac{i}{2}  [r-r_0
\delta_{\alpha\beta}]
\ed
\bd
\hat{R}_\alpha=i\rho ~~~~~~~~~~~ \hat{Q}_\alpha=i\phi
~~~~~~~~~~~
\hat{P}_\alpha(u,v)=i\chi[u,v]
\ed
This allows us to expand
the quantity $\Psi$ of (\ref{eq:finalpsi}) for
small $n$:
\bd
\lim_{n\to 0}\frac{1}{n}\Psi[\ldots]=
-\phi(1\minus q_0)-\rho R
+\frac{1}{2}qr-\frac{1}{2}q_0(r\minus r_0)
-\frac{1}{2}\log r_0 + \frac{1}{2r_0}(r \plus \rho^2 Q)
\ed
\bd
-\int\! dx dy~\chi[x,y] \ P[x,y]
+\lim_{n\to 0}\frac{\alpha}{n} \log \cD[0,0]
+ {\rm constants}
\ed
At this stage it is useful to work out those saddle-point
equations that follow upon variation of $\{\phi,r,\rho,r_0\}$:
\bd
q_0=1
~~~~~~~~~~
r_0=\frac{1}{1-q}
~~~~~~~~~~
\rho=\frac{R}{Q(1\minus q)}
~~~~~~~~~~
r=\frac{qQ\minus R^2}{Q(1\minus
q)^2}
\ed
These allow us to eliminate most variational parameters,
leaving a
saddle-point problem involving only the function
$\chi[x,y]$ and the
scalar $q$:
\bd
\lim_{n\to 0} \ \frac{1}{n} \Psi[q,\{\chi\}]
=
\frac{1\minus R^2/Q}{2(1\minus q)} +\frac{1}{2}
\log(1\minus q)
-\int\! dx dy~\chi[x,y] \ P[x,y]
\ed
\be
+\lim_{n\to 0} \ \frac{\alpha}{n} \log \cD[0,0;q,\{\chi\}]
+ {\rm constants}
\label{eq:simpleRSsaddle}
\ee
Finally we have to work out the RS version of
$\cD[0,0;q,\{\chi\}]$,
as defined more generally in
(\ref{eq:Dasintegral}).
The inverse of the matrix in
(\ref{eq:spinglassops}), in RS ansatz, is found to be:
\be
\bA=\left(
\!\!\begin{array}{cccc} C_{11} & \!\cdots\! & C_{1n}
& \gamma\\
\vdots && \vdots & \vdots \\
C_{n1} & \!\cdots\! & C_{nn} & \gamma \\
\gamma & \!\cdots\! & \gamma & b
\end{array}\!\!\right)
~~~~~~~
C_{\alpha\beta}=\frac{\delta_{\alpha\beta}}{1\minus q}
\ \minus d
~~~~~~~
\begin{array}{l}
\gamma=-\frac{R/\sqrt{Q}}{1- q}+\order(n)
\\[2mm]
b=1+\order(n)
\\[1mm]
d=\frac{q- R^2/Q}{(1- q)^2}+\order(n)
\end{array}
\label{eq:RSinverse}
\ee
With this expression we obtain
\bd
\cD[0,0;q,\{\chi\}]=
\frac{
\int\!d\bx \ dy~
e^{-\frac{1}{2}\bx\cdot\bC\bx-\frac{1}{2}by^2
-\gamma y\sum_{\alpha} x_\alpha
+\frac{1}{\alpha}\sum_\alpha\chi(\sqrt{Q}x_\alpha,y)}}
{\int\!d\bx \ dy~
e^{-\frac{1}{2}\bx\cdot\bC\bx-\frac{1}{2}by^2
-\gamma y\sum_{\alpha}
x_\alpha}}
\ed
\bd
=
\frac{\int\!Dz  Dy\left[
\int\!dx~
e^{
-\frac{x^2}{2(1-q)}+ [z\sqrt{d}-\gamma \frac{y}{\sqrt{b}}
]x
+\frac{1}{\alpha}\chi(\sqrt{Q}x,\frac{y}{\sqrt{b}})}
\right]^n}
{\int\!Dz  Dy \left[\int\!dx~
e^{-\frac{1}{2(1-q)}x^2 +
[z\sqrt{d}-\gamma\frac{y}{\sqrt{b}}]x}\right]^n}
\ed
\bd
\lim_{n\to 0}\frac{\alpha}{n}\log \cD[0,0;q,\{\chi\}]
=
\alpha\int\!Dz  Dy \log \left\{
\frac{\int\!dx~
e^{
-\frac{x^2}{2Q(1-q)}+ x[z\sqrt{d}-\gamma y]/\sqrt{Q}
+\frac{1}{\alpha}\chi(x,y)}}
{\int\!dx~
e^{-\frac{x^2}{2Q(1-q)} + x[z\sqrt{d}-\gamma y]/\sqrt{Q}}}
\right\}
\ed
We can simplify this result by defining
\be
A=R/Q(1\minus q)~~~~~~~~~~B=\sqrt{qQ\minus R^2}/Q(1\minus q)
\label{eq:AandB}
\ee
which gives
\bd
\lim_{n\to 0}\frac{\alpha}{n}\log \cD[0,0;q,\{\chi\}]
=
\alpha\int\!Dz Dy \ \log \left\{
\frac{\int\!dx~
e^{
-\frac{x^2}{2Q(1-q)}+ x[Ay+Bz]
+\frac{1}{\alpha}\chi(x,y)}}
{\int\!dx~
e^{-\frac{x^2}{2Q(1-q)} + x[Ay+Bz]}}
\right\}
\ed
Upon carrying out the $x$-integration in the denominator
of this expression
we can now write (\ref{eq:simpleRSsaddle}) in a surprisingly
simple
form (with the short-hand (\ref{eq:AandB})):
\bd
\lim_{n\to 0}\frac{1}{n}\Psi[q,\{\chi\}]
=
\frac{1\minus\alpha\minus R^2/Q}{2(1\minus q)} +
\frac{1}{2}(1\minus \alpha)\log(1\minus q)
-\int dx
dy~\chi[x,y] \ P[x,y]
\ed
\be
+~\alpha
\int\!Dz  Dy ~\log \int\!dx~
e^{
-\frac{x^2}{2Q(1-q)}+ x[Ay+Bz]
+\frac{1}{\alpha}\chi[x,y]}
\label{eq:simplestRSsaddle}
\ee
Note that (\ref{eq:simplestRSsaddle}) is to be {\em minimised},
both with
respect to $q$ (which originated as an $n(n\minus 1)$-fold
entry in a
matrix, leading to curvature sign change for $n<1$) and
with respect
to $\chi[x,y]$ (obtained from the $n$-fold occurrence of
the function
$\hat{P}[x,y]$, multiplied by $i$, which also leads to
a curvature
sign change).

The remaining saddle point equations correspond to the
(functional)
variation with respect to $\chi$:
\be
{\rm for~all~}x,y:~~~~~~
P[x,y]=\frac{e^{-\frac{1}{2}y^2}}{\sqrt{2\pi}}
\int\!Dz \left\{
\frac{ e^{-\frac{x^2}{2Q(1-q)}+
x[Ay+Bz]+\frac{1}{\alpha}\chi[x,y]}}
{\int\!dx^\prime~ e^{-\frac{x^{\prime 2}}{2Q(1-q)}+
x^\prime[Ay+Bz]+\frac{1}{\alpha}\chi[x^\prime,y]}}
\right\} \ ,
\label{eq:RSsaddle1}
\ee
and $q$  (using equation (\ref{eq:RSsaddle1}) wherever possible):
\bd
\int\!dx dy~P[x,y] \ x^2
-2R\int\!dx  dy~P[x,y] \ x  y
-qQ(\alpha^{-1}\minus 1)+ R^2\alpha^{-1}
\ed
\be
=\left[2\sqrt{qQ\minus R^2}
+\frac{Q(1\minus q)}{\sqrt{qQ\minus R^2}}\right]
\int\!Dz  Dy~ \frac{\partial}{\partial z} \left\{
\frac{\int\!dx~ x \ e^{-\frac{x^2}{2Q(1-q)}+ x[Ay+Bz]
+\frac{1}{\alpha}\chi[x,y]}}
{\int\!dx~ e^{
-\frac{x^2}{2Q(1-q)}+ x[Ay+Bz]
+\frac{1}{\alpha}\chi[x,y]}}\right\}
\label{eq:RSsaddle2}
\ee

\subsection{Explicit Expression for the Green's Function}

In order to work out the Green's function (\ref{eq:finalA})
we
need $\cL[u,v;u^\prime,v^\prime]$ as defined in
(\ref{eq:nastyterm}) which, in turn,
is given in terms of the integrals
(\ref{eq:Dasintegral},\ref{eq:Fasintegral}).
First we calculate in RS ansatz the $n\to 0$ limit of
$D[u,v;q,\{\chi\}]$
(\ref{eq:Dasintegral}), using (\ref{eq:RSinverse}),
and simplify the result with the saddle-point
equation (\ref{eq:RSsaddle1}):
\bd
\lim_{n\to 0}\cD[u,v;q,\{\chi\}]
=\int\!Dz  Dy ~e^{-ivy} \
\frac{\int\!dx~ e^{
-\frac{x^2}{2Q(1-q)}+ x[Ay+Bz]
+\frac{1}{\alpha}\chi[x,y] -iux}}
{\int\!dx~ e^{
-\frac{x^2}{2Q(1-q)}+ x[Ay+Bz]
+\frac{1}{\alpha}\chi[x,y]}}
\ed
\be
=\int\!dx  dy~ P[x,y] \ e^{-ivy-iux}
\label{eq:finalDuv}
\ee
Next we work out $F_\lambda^\alpha[u,v]$
(\ref{eq:Fasintegral})
in RS ansatz,  using (\ref{eq:RSinverse}), with
$\lambda\in\{1,2\}$:
\bd
\lim_{n\to 0}\cF_\lambda^\alpha[u,v]=i\lim_{n\to 0}
~~~~~~~~~~~~~~~~~~~~~~~~~~~~~~~~~~~~~~~~~~~~~~~~~~~~~
~~~~~~~~~~~~~~~~
\vspace*{-3mm}
\ed
\bd
\int\!
Dy Dz~e^{-ivy}\int\!d\bx~e^{\sum_\beta
\left[-\frac{1}{2}\frac{x^2_\beta}{1\minus q}+
[z\sqrt{d}-\gamma y]x_\beta
+\frac{1}{\alpha}\chi[\sqrt{Q}x_\beta,y]\right]
-iux_1\sqrt{Q}}
\partial_\lambda\chi[\sqrt{Q}x_\alpha,y]
\ed
Replica permutation symmetries allow us to
simplify this expression:
\be
\lim_{n\to 0}\cF_\lambda^\alpha[u,v]=\delta_{\alpha
1}  F_\lambda^1[u,v] + (1\minus \delta_{\alpha 1})
 F_\lambda^2[u,v]
\label{eq:bigF}
\ee
with
\be
F_\lambda^1[u,v]
=i\int\!dx \ dy~P[x,y] \ e^{-ivy-iux} \
\partial_\lambda\chi[x,y]
\label{eq:Fsame}
\ee
and
\bd
F_\lambda^2[u,v]=
i\int\!
Dy  Dz~e^{-ivy}
~~~~~~~~~~~~~~~~~~~~~~~~~~~~~~~~~~~~~~~~~
~~~~~~~~~~~~~~~~~~~~~~~~~~~~~
\vspace*{-3mm}
\ed
\be
\frac{
\left[\int\!dx~e^{-\frac{x^2}{2Q(1\minus
q)}+x[Ay+Bz]+\frac{1}{\alpha}\chi[x,y]}\
\partial_\lambda\chi[x,y]\right]
\left[\int\!dx~e^{-\frac{x^2}{2Q(1\minus
q)}+x[Ay+Bz]+\frac{1}{\alpha}\chi[x,y]-iux}
\right]}
{\left[\int\!dx~e^{-\frac{x^2}{2Q(1\minus
q)}+x[Ay+Bz]+\frac{1}{\alpha}\chi[x,y]}\right]^2}
\label{eq:Fdiff}
\ee
We can now proceed with the calculation of
(\ref{eq:nastyterm}),
whose building blocks are
\bd
\alpha^{-1}\cF^\alpha_{1}[u,v]+u\delta_{\alpha 1}\cD[u,v]=
\delta_{\alpha 1}G_{1}[u,v]+(1\minus
\delta_{\alpha 1})\tilde{G}_{1,2}[u,v]
\vspace*{-1mm}
\ed
\bd
\alpha^{-1}\cF^\alpha_{2}[u,v]+v\delta_{\alpha 1}\cD[u,v]=
\delta_{\alpha 1}G_{2}[u,v]+(1\minus
\delta_{\alpha 1})\tilde{G}_{2}[u,v]
\ed
with
\bd
G_{1}[u,v]=\alpha^{-1}\cF^1_{1,2}[u,v]+u\cD[u,v]~~~~~~~~~~
\tilde{G}_1[u,v]=\alpha^{-1}\cF^2_1[u,v]
\vspace*{-1mm}
\ed
\bd
G_2[u,v]=\alpha^{-1}\cF^1_2[u,v]+v\cD[u,v]~~~~~~~~~~
\tilde{G}_2[u,v]=\alpha^{-1}\cF^2_2[u,v]
\ed
and their Fourier transforms:
\bd
\hat{G}_1[\hat{u},\hat{v}]=
\int\!\frac{du \ dv}{(2\pi)^2}e^{iu\hat{u}+iv\hat{v}}G_1[u,v]
~~~~~~~~~~
\hatildeG_1[\hat{u},\hat{v}]=
\int\!\frac{du \ dv}{(2\pi)^2}e^{iu\hat{u}+
iv\hat{v}}\tilde{G}_1[u,v]
\vspace*{-1mm}
\ed
\bd
\hat{G}_2[\hat{u},\hat{v}]=
\int\!\frac{du \ dv}{(2\pi)^2}e^{iu\hat{u}+
iv\hat{v}}G_2[u,v]
~~~~~~~~~~
\hatildeG_2[\hat{u},\hat{v}]=
\int\!\frac{du \ dv}{(2\pi)^2}e^{iu\hat{u}+
iv\hat{v}}\tilde{G}_2[u,v]
\ed
With these short-hands we obtain a compact expression for
(\ref{eq:nastyterm}),
and can subsequently write our expression (\ref{eq:finalA})
for the Green's function
$\A[x,y;x^\prime,y^\prime]$ as
\bd
\A[x,y;x^\prime,y^\prime]
=
-Q(1\minus q)\left[\hat{G}_1[x,y]\hat{G}_1[x^\prime,y^\prime]
-\hatildeG_1[x,y]\hatildeG_1[x^\prime,y^\prime]\right]
\vspace*{-1mm}
\ed
\bd
-Q q
\left[\hat{G}_1[x,y]-\hatildeG_1[x,y]\right]
\left[\hat{G}_1[x^\prime,y^\prime]-
\hatildeG_1[x^\prime,y^\prime]\right]
\ed
\bd
-R
\left[\hat{G}_1[x,y]- \hatildeG_1[x,y]\right]
\left[\hat{G}_2[x^\prime,y^\prime]-\hatildeG_2
[x^\prime,y^\prime]\right]
\ed
\bd
-R
\left[\hat{G}_1[x^\prime,y^\prime]-\hatildeG_1
[x^\prime,y^\prime]\right]
\left[\hat{G}_2[x,y]-\hatildeG_2[x,y]\right]
\ed
\be
-
\left[\hat{G}_2[x,y]-\hatildeG_2[x,y]\right]
\left[\hat{G}_2[x^\prime,y^\prime]-\hatildeG_2
[x^\prime,y^\prime]\right]
\label{eq:RSA}
\ee
Finally, working out the four relevant Fourier transforms,
using (\ref{eq:finalDuv},\ref{eq:Fsame},\ref{eq:Fdiff}),
gives:
\bd
\hat{G}_1[x,y]=
i\left[\frac{1}{\alpha} \ P[x,y] \
\frac{\partial}{\partial x}\chi[x,y]
-\frac{\partial}{\partial x} P[x,y] \right]
\ed
\bd
\hat{G}_2[x,y]=
i\left[\frac{1}{\alpha} \ P[x,y] \
\frac{\partial}{\partial y}\chi[x,y]
-\frac{\partial}{\partial y} P[x,y] \right]
\ed

\clearpage
\bd
\hatildeG_1[x,y]=\frac{i}{\alpha}
\frac{e^{-\frac{1}{2}y^2}}{\sqrt{2\pi}}\int\!Dz
~~~~~~~~~~~~~~~~~~~~~~~~~~~~~~~~~~~~~~~~
~~~~~~~~~~~~~~~~~~~~~~~~~~~~~~
\vspace*{-1mm}
\ed
\bd
\frac{
\left[\int\!dx^\prime~e^{-\frac{x^{\prime 2}}{2Q(1\minus
q)}+x^\prime[Ay+Bz]+\frac{1}{\alpha}\chi[x^\prime,y]} \
\partial_1\chi[x^\prime,y]\right]
e^{-\frac{x^2}{2Q(1\minus
q)}+x[Ay+Bz]+\frac{1}{\alpha}\chi[x,y]}}
{\left[\int\!dx^\prime~e^{-\frac{x^{\prime 2}}{2Q(1\minus
q)}+x^\prime[Ay+Bz]+\frac{1}{\alpha}\chi[x^\prime,y]}\right]^2}
\ed
\bd
\hatildeG_2[x,y]=
\frac{i}{\alpha}
\frac{e^{-\frac{1}{2}y^2}}{\sqrt{2\pi}}\int\!Dz
~~~~~~~~~~~~~~~~~~~~~~~~~~~~~~~~~
~~~~~~~~~~~~~~~~~~~~~~~~~~~~~~~~~~~~~
\vspace*{-1mm}
\ed
\bd
\frac{
\left[\int\!dx^\prime~e^{-\frac{x^{\prime 2}}{2Q(1\minus
q)}+x^\prime[Ay+Bz]+\frac{1}{\alpha}\chi[x^\prime,y]} \
\partial_2\chi[x^\prime,y]\right]
e^{-\frac{x^2}{2Q(1\minus
q)}+x[Ay+Bz]+\frac{1}{\alpha}\chi[x,y]}
}
{\left[\int\!dx^\prime~e^{-\frac{x^{\prime 2}}{2Q(1\minus
q)}+x^\prime[Ay+Bz]+\frac{1}{\alpha}\chi[x^\prime,y]}\right]^2}
\ed

\subsection{Summary}

At this stage it is advantageous to summarize the theory and choose
the most transparent representation of our equations.
We first replace the function  $\chi[x,y]$ by the effective measure
$M[x,y]$:
\bd
M[x,y]=
e^{-\frac{x^{2}}{2Q(1\minus
q)}+Axy+\frac{1}{\alpha}\chi[x^\prime,y]}
\ed
We introduce a compact notation for the various
averages we encounter:
\be
\bra f[x,y,z]\ket_\star =
 \frac{\int\!dx~M[x,y]e^{Bxz}f[x,y,z]}
{\int\!dx~M[x,y]e^{Bxz}}
~~~~~~~~
\bigbras f[y,z]\bigkets=\int\!Dy Dz ~f[y,z]
\label{eq:defaverages}
\ee
with the short-hand
\vspace*{-1mm}
\bd
B=\frac{\sqrt{qQ-R^2}}{Q(1-q)}
\ed
From  equation (\ref{eq:RSsaddle1}) we deduce
that
the function $P[x,y]$ always obeys
\be
P[x,y]=P[x|y] \ P[y] ~~~~~~~~~~
P[y]=(2\pi)^{-\frac{1}{2}}e^{-\frac{1}{2}y^2}
\label{eq:conditionalsummary}
\ee
This enables us to write our results
in terms of $P[x|y]$ rather than $P[x,y]$.
We also introduce the transformed Green's function $\tilde{\A}$
via
$\cA[x,y;x^\prime,y^\prime]=
P[x,y] \ \tilde{\cA}[x,y;x^\prime,y^\prime] \ P[x^\prime,y^\prime]$.
In combination these simplifications
allow us to summarize our theory as follows. The macroscopic
dynamic equations are:
\be
\frac{d}{dt}Q =
2\eta \int\!dx Dy~P[x|y] \ x \ \G[x,y]
+ L
~~~~~~~~~~
\frac{d}{dt}R
=\eta  \int\!dx Dy~ P[x|y] \ y \ \G[x,y]
\label{eq:dQdRdt}
\ee
\bd
\frac{\partial }{\partial t}P[x|y] = \!
-\eta\frac{\partial}{\partial x}\!\left\{P[x|y]\left[
\frac{1}{\alpha}\G[x,y]
+\int\!dx^\prime Dy^\prime P[x^\prime|y^\prime]
\G[x^\prime,y^\prime]  \tilde{\cA}[x,y;x^\prime,y^\prime]
\right]\right\}
\ed
\be
+\frac{1}{2}L\frac{\partial^2}{\partial x^2}P[x|y]
\vspace*{-2mm}
\label{eq:dPdtnewsummary}
\ee
\clearpage
\noindent with
\be
{\rm batch:}~~L=0
~~~~~~~~~~~~
{\rm on\!-\!line:}~~
L=\eta^2\int\!dx Dy~ P[x|y]\G^2[x,y]
\label{eq:L}
\ee
The solution of (\ref{eq:dQdRdt},\ref{eq:dPdtnewsummary})
subsequently generates the training- and generalization
errors (\ref{eq:Et},\ref{eq:Eg}) at any time:
\be
E_{\rm t}=\int\!dx  Dy~P[x|y] \ \theta[-xy]
~~~~~~~~~~
E_{\rm g}=\frac{1}{\pi} \ \arccos[R/\sqrt{Q}]
\label{eq:errors}
\ee
In order to determine the Green's function
$\tilde{\cA}[x,y;x^\prime,y^\prime]$
in equation (\ref{eq:dPdtnewsummary}) one first has to
calculate the auxiliary
order parameters $q$ and $\{M[x,y]\}$ by solving the
following two saddle point equations:
\be
{\rm for~all}~X,y:~~~~~
P[X|y]
=\int\! Dz~ \bra \delta[X\minus x]\ket_\star
\label{eq:saddle1newsummary}
\ee
\be \mbox{\small $
\int\!dx  Dy~P[x|y] \ (x^2\minus 2Rxy)
-qQ  \left(\frac{1}{\alpha}\minus 1\right)+ \frac{R^2}{\alpha}
=\left[2\sqrt{qQ\minus R^2}\plus \frac{1}{B}\right]
\int\!Dy  Dz~ z\bra x\ket_\star $}
\label{eq:saddle2newsummary}
\ee
The Green's function in equation (\ref{eq:dPdtnewsummary})
is then given by
\bd
\tilde{\A}[x,y;x^\prime,y^\prime]=
Q(1\minus q)\left[J_1[x,y]J_1[x^\prime,y^\prime]\minus
\tilde{J}_1[x,y]\tilde{J}_1[x^\prime,y^\prime]\right]
\vspace*{-2mm}
\ed
\bd
+Q q
\left[J_1[x,y]\minus \tilde{J}_1[x,y]\right]
\left[J_1[x^\prime,y^\prime]\minus \tilde{J}_1
[x^\prime,y^\prime]\right]
+J_2[x,y]J_2[x^\prime,y^\prime]
\ed
\be
+R
\left[J_1[x,y]\minus  \tilde{J}_1[x,y]\right]J_2
[x^\prime,y^\prime]
+R
\left[J_1[x^\prime,y^\prime]\minus \tilde{J}_1
[x^\prime,y^\prime]\right]
J_2[x,y]
\label{eq:RSAnewsummary}
\ee
with the functions
\be
J_1[X,Y]=
\frac{\partial}{\partial X}\log\frac{ M[X,Y]}{P[X|Y]}
+\frac{X\minus RY}{Q(1\minus q)}
~~~~~~~~~~~~~~~~~~~~~~~~~~~~~~~~~~~~~
\label{eq:J1}
\ee
\be
\tilde{J}_1[X,Y]=P[X|Y]^{-1}\int\!Dz
\bigbra
\frac{\partial}{\partial x}\log M[x,Y]+
\frac{x\minus RY}{Q(1\minus q)}
\bigket_\star
\bra \delta[X\minus x]\ket_\star
\label{eq:J1tilde}
\ee
\bd
J_2[X,Y]=\frac{\partial}{\partial Y}\log
\frac{M[X,Y]}{P[X|Y]}
-\frac{RX}{Q(1\minus q)}+Y
~~~~~~~~~~~~~~~~~~~~~~~~~~~~~~~~~~~~~~~~
\ed
\be
-P[X|Y]^{-1}\int\!Dz
\bigbra
\frac{\partial}{\partial Y}\log M[x,Y]-
\frac{Rx}{Q(1\minus q)}
\bigket_\star
\bra \delta[X\minus x]\ket_\star
\label{eq:J2}
\ee
We finally
work out a simple inequality to determine for
which regime of $q$-values one should inspect the
saddle-point equations.
From (\ref{eq:interpretation}) it follows that
$q=\bra\frac{1}{N}\sum_i \bra \sigma_i\ket^2\ket_{\ \rmsets}$,
giving
\bd
0\leq \bigbra\frac{1}{N}\sum_i\left[\bra
\sigma_i\ket-\tau_i\left[\frac{1}{N}\sum_j\tau_j\bra
\sigma_j\ket\right]\right]^2\bigket_{\rmsets}
=q-R^2/Q
\ed

\section{Applications and Tests of the Theory}

\subsection{The Limit $\alpha\to\infty$}

The very least we should require of our theory  is that it reduces
to the
simple $(Q,R)$ formalism of infinite training sets in the limit
$\alpha\to\infty$. This indeed happens.
For $\alpha\to\infty$ our macroscopic equations
(\ref{eq:dQdRdt},\ref{eq:dPdtnewsummary})
reduce to
\be
\frac{d}{dt}Q =
2\eta \int\!dx  Dy~P[x|y] \ x \ \G[x;y]
+ L
~~~~~~~~
\frac{d}{dt}R
=\eta  \int\!dx Dy~ P[x|y]\ y \ \G[x;y]
\label{eq:largealphaQR}
\ee
\be
\frac{\partial }{\partial t}P[x|y] =
-\eta\frac{\partial}{\partial x}\!\left\{\!
P[x|y]
\int\!dx^\prime Dy^\prime P[x^\prime|y^\prime] \
\G[x^\prime,y^\prime] \tilde{\cA}[x,y;x^\prime,y^\prime]
\right\}
+\frac{L}{2}\frac{\partial^2}{\partial x^2}P[x|y]
\label{eq:largealphaP}
\ee
with the Green's function $\tilde{\cA}
[x,y;x^\prime,y^\prime]$
(\ref{eq:RSAnewsummary}), and with $L$ as given in
(\ref{eq:L}).
The saddle-point equations from which to solve $\{M[x,y]\}$
and $q$ are:
\bd
P[X|y]
=\int\! Dz~ \bra \delta[X\minus x]\ket_\star
\ed
\bd
qQ+\int\!dx \ Dy~P[x|y] \ (x^2\minus 2Rxy)
=\left[2\sqrt{qQ\minus R^2}\plus B^{-1}\right]
\int\!Dy  Dz~ z\bra x\ket_\star
\ed
with the convention (\ref{eq:defaverages}),
and with the short-hand
$B=\sqrt{qQ\minus R^2}/Q(1\minus q)$.
We now make the following ansatz:
\be
P[x|y]=[2\pi(Q\minus R^2)]^{-\frac{1}{2}}~
e^{-\frac{1}{2}[x-Ry]^2/(Q-R^2)}
\label{eq:ansatz}
\ee
and find that the two saddle-point equations are
simultaneously solved by
\be
M[x,y]=[2\pi Q(1\minus q)]^{-\frac{1}{2}}~
e^{-\frac{1}{2}[x-Ry]^2/Q(1-q)}
\label{eq:largealphaM}
\ee
(we will return to the question of uniqueness later). The values of
$Q$ and $R$ at any time are fully specified by
(\ref{eq:largealphaQR}); what remains is to verify the validity of
(\ref{eq:largealphaP}).
Given the measure (\ref{eq:largealphaM}) one can calculate
the three functions $J_1[x,y]$,  $\tilde{J}_1[x,y]$  and
$J_2[x,y]$, and thus the Green's function (\ref{eq:RSAnewsummary}),
explicitly:
\bd
J_1[x,y]=\frac{x\minus Ry}{Q\minus R^2}
~~~~~~~~~~
\tilde{J}_1[x,y]=0
~~~~~~~~~~
J_2[x,y]=\frac{Qy\minus Rx}{Q\minus R^2}
\ed
\be
\tilde{\A}[x,y;x^\prime,y^\prime]
=\frac{(x\minus Ry)x^\prime+(Qy\minus Rx)y^\prime}{Q\minus R^2}
\ee
This, in combination with the equations
(\ref{eq:largealphaQR}) for $Q$ and $R$,
leads to an explicit expression for the diffusion
equation (\ref{eq:largealphaP}):
\bd
\frac{\partial }{\partial t}P[x|y]
=\frac{\partial}{\partial x}\left\{
\frac{L}{2}\frac{\partial}{\partial x}P[x|y]
-\frac{P[x|y]}{Q\minus R^2}\left[
\frac{1}{2}(x\minus Ry)[\frac{dQ}{dt}-L]
+(Qy\minus Rx)\frac{dR}{dt}
\right]\right\}
\ed
Insertion of our ansatz (\ref{eq:ansatz}) into both sides of
this
equation, followed by some rearranging of terms,
 shows that it is indeed satisfied.
This confirms that from our general theory
we indeed recover for $\alpha\to\infty$ the standard theory
for infinite training sets, i.e. the closed set
(\ref{eq:largealphaQR},\ref{eq:ansatz}), as claimed.

\subsection{Locally Gaussian Solutions}

As soon as $\alpha$ is finite the diffusion equation
(\ref{eq:dPdtnewsummary}) will no longer have Gaussian
solutions.
However, we will show in this section that
for a specific simple class of learning rules one can
find solutions
described by a conditional distribution $P[x|y]$ which
for each $y$ is
Gaussian in
$x$, but with moments which are non-trivial functions
of $y$ (such
that the full distribution $P[x,y]$ is not Gaussian):
\be
\G[x,y]=\G_0[y]+x\G_1[y]:
~~~~~~~~~~
P[x|y]=\frac{e^{-\frac{1}{2}
[x-\overline{x}(y)]^2/\Delta^2(y)}}{\Delta(y)\sqrt{2\pi}}
\label{eq:localgauss}
\ee
We choose the distribution $P[x|y]$ in
(\ref{eq:localgauss}) as our ansatz.
Insertion into the saddle-point equation
(\ref{eq:saddle1newsummary}) and integration over $z$
shows that (\ref{eq:saddle1newsummary}) is solved by the
following measure (the issue of uniqueness will be
discussed later):
\be
M[x,y]=
\frac{e^{-\frac{1}{2}
[x-\overline{x}(y)]^2/\sigma^2(y)}}{\sigma(y)\sqrt{2\pi}}
\label{eq:localgaussmeasure}
\ee
with the usual short-hand
$B=\sqrt{qQ\minus R^2}/Q(1\minus q)$, and with
\be
\sigma^2(y)=\frac{1}{2B^2}[\sqrt{1+4B^2\Delta^2(y)}-1]
\label{eq:localsigma}
\ee
The simple form of (\ref{eq:localgaussmeasure}) allows us to
proceed analytically, using identities such as
\bd
\bra x\ket_\star
=\overline{x}(y)+Bz\sigma^2(y)
~~~~~~~~~~
\bra [x-\overline{x}(y)]^2\ket_\star=
\sigma^2(y)+B^2z^2\sigma^4(y)
\ed
The remaining saddle point equation
(\ref{eq:saddle2newsummary})
 for $q$ can be simplified to
\be
\int\!Dy\left[\Delta^2(y)\plus \overline{x}^2(y)\minus
2Ry \ \overline{x}(y)\right]\!
-qQ\left(\frac{1}{\alpha}\minus 1\!\right)+ \frac{R^2}{\alpha}
=\!\left[\frac{2(qQ\minus R^2)}{Q(1\minus q)}\plus 1\right]\!
\int\!Dy~\sigma^2(y)
\label{eq:localgausssaddle2}
\ee
Next we have to show that our ansatz (\ref{eq:localgauss}), in
combination with (\ref{eq:localsigma},\ref{eq:localgausssaddle2}),
solves the dynamic equations
(\ref{eq:dQdRdt},\ref{eq:dPdtnewsummary}).
In order to do so we first calculate the building blocks of the
Green's function $\tilde{A}[x,y;x^\prime,y^\prime]$, giving
(after a modest amount of bookkeeping):
\bd
J_1[x,y]=xV_1(y)+V_2(y)
~~~~~~
\tilde{J}_1[x,y]=x\tilde{V}_1(y)+\tilde{V}_2(y)
~~~~~~
J_2[x,y]=xW_1(y)+W_2(y)
\ed
with the six functions
\bd
V_1(y)=\frac{\sigma^2(y)\Delta^2(y)\plus Q(1\minus
q) \ [\sigma^2(y)\minus\Delta^2(y)]}
{Q(1\minus q) \ \sigma^2(y) \ \Delta^2(y)}
\ed
\bd
V_2(y)=\frac{\overline{x}(y) \ Q(1\minus
q) \ [\Delta^2(y)\minus\sigma^2(y)]\minus Ry \ \sigma^2(y)
\ \Delta^2(y)}
{Q(1\minus q) \ \sigma^2(y) \ \Delta^2(y)}
\ed
\bd
\tilde{V}_1(y)=
\frac{[\sigma^2(y)\minus \Delta^2(y)] \ [Q(1\minus q)\minus
\sigma^2(y)]}
{Q(1\minus q) \ \sigma^2(y) \ \Delta^2(y)}
\ed
\bd
\tilde{V}_2(y)=
\frac{\overline{x}(y)[\Delta^2(y)\minus \sigma^2(y)] \
[Q(1\minus q)\minus \sigma^2(y)]+(\overline{x}(y) \minus Ry)
\ \Delta^2(y) \ \sigma^2(y)}
{Q(1\minus q) \ \sigma^2(y) \ \Delta^2(y)}
\ed
\bd
W_1(y)= -
\frac{R\sigma^4(y)}{Q(1\minus q) \ \sigma^2(y) \ \Delta^2(y)}
\ed
\bd
W_2(y)=
\frac{y \ \sigma^2(y) \ \Delta^2(y) \
Q(1\minus q)\plus R \ \overline{x}(y) \ \sigma^4(y)]}
{Q(1\minus q) \ \sigma^2(y) \ \Delta^2(y)}
\ed
Insertion of these expressions into the Green's function
(\ref{eq:RSAnewsummary}) subsequently gives the simple
expression
\be
\tilde{\A}[x,y;x^\prime,y^\prime]
=x x^\prime U_1(y,y^\prime)+x U_2(y,y^\prime) +x^\prime
U_2(y^\prime,y) +U_3(y,y^\prime)
\label{eq:localgaussgreen}
\ee
with the three kernels
%%%%%%%%%%%%%%%%%%%%%%%% spaces
\bd
U_1(y,y^\prime)=W_1(y)W_1(y^\prime)
~~~~~~~~~~~~~~~~~~~~~~~~~~~~~~~~~~~~~~~~~~~~~
~~~~~~~~~~~~~~~~~~~~~~~~~~~~~~~
\ed
\bd
+
Q(1\minus q)\left[V_1(y)V_1(y^\prime)\minus
\tilde{V}_1(y)\tilde{V}_1(y^\prime)\right]
+Qq\left[V_1(y)\minus \tilde{V}_1(y)\right]
\left[V_1(y^\prime)\minus \tilde{V}_1(y^\prime)\right]
\ed
\be
+R \ [V_1(y)\minus
\tilde{V}_1(y)] \ W_1(y^\prime)+R \ W_1(y)
\ [V_1(y^\prime)\minus \tilde{V}_1(y^\prime)]
\label{eq:U1}
\ee
\bd
U_2(y,y^\prime)=W_1(y)W_2(y^\prime)
~~~~~~~~~~~~~~~~~~~~~~~~~~~~~~~~~~~~~~~~~~~
~~~~~~~~~~~~~~~~~~~~~~~~~~~~~~~~~
\ed
\bd
+
Q(1\minus q)\left[V_1(y)V_2(y^\prime)\minus
\tilde{V}_1(y)\tilde{V}_2(y^\prime)\right]
+Qq\left[V_1(y)\minus \tilde{V}_1(y)\right]
\left[V_2(y^\prime)\minus
\tilde{V}_2(y^\prime)\right]
\ed
\be
+R \ [V_1(y)\minus \tilde{V}_1(y)] \ W_2(y^\prime)
+R \ W_1(y) \ [V_2(y^\prime)\minus \tilde{V}_2(y^\prime)]
\label{eq:U2}
\ee
\bd
U_3(y,y^\prime)=W_2(y)W_2(y^\prime)
~~~~~~~~~~~~~~~~~~~~~~~~~~~~~~~~~~~~~~~~~~~~~~~~~~~~~~~
~~~~~~~~~~~~~~~~~~~~~
\ed
\bd
+
Q(1\minus q)\left[V_2(y)V_2(y^\prime)\minus
\tilde{V}_2(y)\tilde{V}_2(y^\prime)\right]
+Q q\left[V_2(y)\minus \tilde{V}_2(y)\right]
\left[V_2(y^\prime)\minus
\tilde{V}_2(y^\prime)\right]
\ed
\be
+R \ [V_2(y)\minus \tilde{V}_2(y)] \ W_2(y^\prime)
+R \ W_2(y) \ [V_2(y^\prime)\minus \tilde{V}_2(y^\prime)]
\label{eq:U3}
\ee
Insertion of the above expression (\ref{eq:localgaussgreen})
for the Green's function
into the right-hand side of our diffusion equation
(\ref{eq:dPdtnewsummary})
gives, in turn:
\bd
\frac{\Delta^2(y)}{P[x|y]}~{\rm RHS}
=-\frac{L}{2}\left[
1-\frac{[x\minus \overline{x}(y)]^2}{\Delta^2(y)}\right]
+\eta[x\minus \overline{x}(y)]\left\{
\frac{1}{\alpha}\G[x,y]
~~~~~~~~~~~~~~~~
\right.
\ed
\bd
\left.
+[x\minus \overline{x}(y)]
\int\!dx^\prime Dy^\prime~P[x^\prime|y^\prime] \
\G[x^\prime,y^\prime] \
\left[x^\prime U_1(y,y^\prime) \plus
U_2(y,y^\prime)\right]
\right.
\ed
\bd
\left.
+\overline{x}(y)
\int\!dx^\prime Dy^\prime~P[x^\prime|y^\prime]
\ \G[x^\prime,y^\prime] \
\left[x^\prime U_1(y,y^\prime) \plus U_2(y,y^\prime)\right]
\right.
\ed
\bd
\left.
+\int\!dx^\prime Dy^\prime~P[x^\prime|y^\prime]
\ \G[x^\prime,y^\prime] \
\left[x^\prime U_2(y^\prime,y)\plus U_3(y,y^\prime)\right]
\right\}
\ed
\bd
~~~~~~~~~~
-\eta\Delta^2(y)\left\{
\frac{1}{\alpha}\frac{\partial}{\partial x}\G[x,y]
+\!\int\!dx^\prime Dy^\prime~P[x^\prime|y^\prime]
\ \G[x^\prime,y^\prime]
\left[x^\prime U_1(y,y^\prime) \plus U_2(y,y^\prime)\right]
\right\}
\ed
For the left-hand side of the diffusion equation for
$P[x|y]$ we
get
\bd
\frac{\Delta^2(y)}{P[x|y]}~{\rm LHS}
=[x\minus\overline{x}(y)]\frac{d}{dt}\overline{x}(y)
+\left[\frac{[x\minus
\overline{x}(y)]^2}{\Delta(y)}-\Delta(y)\right]
\frac{d}{dt}\Delta(y)
\ed
Since LHS is a second-order polynomial in $x$, we
have to restrict ourselves
to functions $\G[x,y]$ which are
first-order polynomials in $x$, hence the definition
(\ref{eq:localgauss}).
We equate the three monomials of
$x\minus\overline{x}(y)$ and find for self-consistent
solutions
of the locally Gaussian type (\ref{eq:localgauss})
the following conditions:
\bd
[x-\overline{x}(y)]^2:~~~~~
\frac{d}{dt}\Delta(y)
=\frac{L}{2\Delta(y)}
+\eta\Delta(y)\left\{\alpha^{-1}\G_1(y)
~~~~~~~~~~~~~~~~~~~~
\room
\right.
\ed
\bd
\left.
~~~~~~~~~~~~~~~~~~~~~~~~~~~~~~
+\int\!dx^\prime Dy^\prime~P[x^\prime|y^\prime]
\ \G[x^\prime,y^\prime]
\left[x^\prime U_1(y,y^\prime) + U_2(y,y^\prime)\right]
\right\}
\ed
\bd
[x-\overline{x}(y)]:~~~~~
\frac{d}{dt}\overline{x}(y)
=
\eta\left\{
\alpha^{-1}
\left[\G_0(y)+\overline{x}(y) \ \G_1(y)\right]
\room
~~~~~~~~~~~~~~~~~~~~~
\right.
\ed
\bd
\left.
~~~~~~~~~~~~~~~~~~~~~~~~~
+\overline{x}(y)
\int\!dx^\prime Dy^\prime~P[x^\prime|y^\prime]
\ \G[x^\prime,y^\prime]
\left[x^\prime U_1(y,y^\prime) + U_2(y,y^\prime)\right]
\right.
\ed
\bd
\left.
~~~~~~~~~~~~~~~~~~~~~~~~~~~~~~
+\int\!dx^\prime Dy^\prime~P[x^\prime|y^\prime]
\  \G[x^\prime,y^\prime]
\left[x^\prime U_2(y^\prime,y)+ U_3(y,y^\prime)\right]
\right\}
\ed
\bd
[x-\overline{x}(y)]^0:
~~~~~
\frac{d}{dt}\Delta(y)
=\frac{L}{2\Delta(y)}
+\eta\Delta(y)\left\{\alpha^{-1}\G_1(y)
~~~~~~~~~~~~~~~~~~~~
\room
\right.
\ed
\bd
\left.
~~~~~~~~~~~~~~~~~~~~~~~~~~~~~~
+\int\!dx^\prime Dy^\prime~P[x^\prime|y^\prime]
\ \G[x^\prime,y^\prime]
\left[x^\prime U_1(y,y^\prime) + U_2(y,y^\prime)\right]
\right\}
\ed
This is an important self-consistency test, since
we have three
equations for only two functions
($\overline{x}(y)$ and $\Delta(y)$). Two of the three
equations are found to be identical. In the remaining
two equations
we insert the form of the learning rules (\ref{eq:localgauss})
and perform the
$x$-integrations, giving
\bd
\frac{d}{dt}\Delta(y)
=\frac{L}{2\Delta(y)}
+\eta\Delta(y)\left\{
\frac{\G_1(y)}{\alpha}
+\int\!Dy^\prime~\G_0(y^\prime)
\left[\overline{x}(y^\prime)  \ U_1(y,y^\prime)
\plus U_2(y,y^\prime)\right]
\right.
\ed
\be
\left.
+\int\!Dy^\prime~\G_1(y^\prime)
\left[[\overline{x}^2(y^\prime)\plus\Delta^2(y^\prime)]
\  U_1(y,y^\prime)
+ \overline{x}(y^\prime) \ U_2(y,y^\prime)\right]
\right\}
\label{eq:localvariance}
\ee
\bd
\frac{d}{dt}\overline{x}(y)
=
\eta\left\{
\frac{1}{\alpha}
\left[\G_0(y)\plus\overline{x}(y) \ \G_1(y)\right]
+\overline{x}(y)
\!\int\!Dy^\prime~\G_0(y^\prime)
\left[\overline{x}(y^\prime) U_1(y,y^\prime) \plus
U_2(y,y^\prime)\right]
\right.
\ed
\bd
\left.
+\overline{x}(y)
\!\int\!Dy^\prime~\G_1(y^\prime)
\left[[\overline{x}^2(y^\prime)\plus\Delta^2(y^\prime)]
\ U_1(y,y^\prime)
+ \overline{x}(y^\prime) \ U_2(y,y^\prime)\right]
\right.
\ed
\bd
\left.
+\int\!Dy^\prime~ \G_0(y^\prime)
\left[\overline{x}(y^\prime)  \ U_2(y^\prime,y)
\plus U_3(y,y^\prime)\right]
\right.
\ed
\be
\left.
+\int\!Dy^\prime~\G_1(y^\prime)
\left[[\overline{x}^2(y^\prime)\plus\Delta^2(y^\prime)]
\ U_2(y^\prime,y)
+\overline{x}(y^\prime)  \ U_3(y,y^\prime)\right]
\right\}
\label{eq:localaverage}
\ee
Our result is a solution in the form of
coupled equations (\ref{eq:dQdRdt},\ref{eq:localgauss},
\ref{eq:localgausssaddle2},\ref{eq:localvariance},
\ref{eq:localaverage}),
without a functional saddle-point problem and without
a diffusion
equation.

\subsection{Explicit Example: Hebbian Learning}

The simplest non-trivial member of the family
(\ref{eq:localgauss}) of learning rules with locally
Gaussian
field distributions is the Hebb rule: $\G[x,y]=\sgn[y]$
(i.e. $\G_0(y)=\sgn(y)$ and $\G_1(y)=0$).
This example we will work out in full, as an explicit
illustration and in order to have a precise test (since only
for this rule one can calculate the macroscopic observables
exactly and directly from the microscopic laws).
The remaining integrals in our equations
(\ref{eq:dQdRdt},\ref{eq:localgauss},
\ref{eq:localgausssaddle2},\ref{eq:localvariance},
\ref{eq:localaverage})
can be carried out explicitly. We find
(for initial states $\bJ(0)$ which are not correlated
with the
questions in the training set):
\be
R=R_0+\eta t\sqrt{\frac{2}{\pi}}
~~~~~~~~
Q=Q_0+\eta t\left[\kappa\eta\plus 2R_0\sqrt{\frac{2}{\pi}}
\right]
+\eta^2 t^2\left[\frac{2}{\pi}\plus \frac{1}{\alpha}\right]
\label{eq:HebbQR}
\ee
(with $\kappa=1$ for on-line learning and $\kappa=0$ for
batch
learning),
and
\be
\overline{x}(y)=Ry+\alpha^{-1}\eta t\sgn(y)
\label{eq:Hebbaverage}
\ee
To find the width $\Delta$ of $P[x|y]$ (which is
found to remain independent of $y$, if so at $t=0$) , we
have to solve the coupled equations
(\ref{eq:localgausssaddle2},\ref{eq:localvariance}),
which for the
Hebb rule reduce to
\bd
\frac{1}{\eta}\frac{d}{dt}\Delta^2
=\kappa\eta
+\frac{\eta t}{\alpha} \frac{Q(1\minus q)}
{qQ\minus R^2}
\left[\sqrt{1+\frac{4(qQ\minus R^2)\Delta^2}{Q^2(1\minus q)^2}}
-1\right]
\ed
\bd
%%%%%%%%%%%%%%% minor modification
(R^2\minus qQ)\left(\frac{1}{\alpha}\minus 1\right)
+\Delta^2+\frac{\eta^2t^2}{\alpha^2} \!
=Q(1\minus q)\left[1\plus \frac{Q(1\minus q)}{2(qQ\minus R^2)}
\right]
\left[\sqrt{1\plus \frac{4(qQ\minus R^2)\Delta^2}{Q^2(1\minus q)^2}}
\minus 1\right]
\ed
The solution of these equations is given by
\be
\Delta^2=Q-R^2
~~~~~~~~~~
q=Q^{-1}[R^2+\alpha^{-1}\eta^2 t^2]
\label{eq:HebbDq}
\ee
This solution is unique (for a proof see Coolen and Saad,
1998).
Thus:
\be
P[x|y]=\frac{e^{-\frac{1}{2}
[x-Ry-\alpha^{-1}\eta t\sgn(y)]^2/(Q-R^2)}}
{\sqrt{2\pi(Q-R^2)}}
\label{eq:HebbP}
\ee
We can now also calculate both errors as a functions of time:
\begin{figure}[t]
\centering
\vspace*{87mm}
\hbox to \hsize{\hspace*{-1mm}\includegraphics{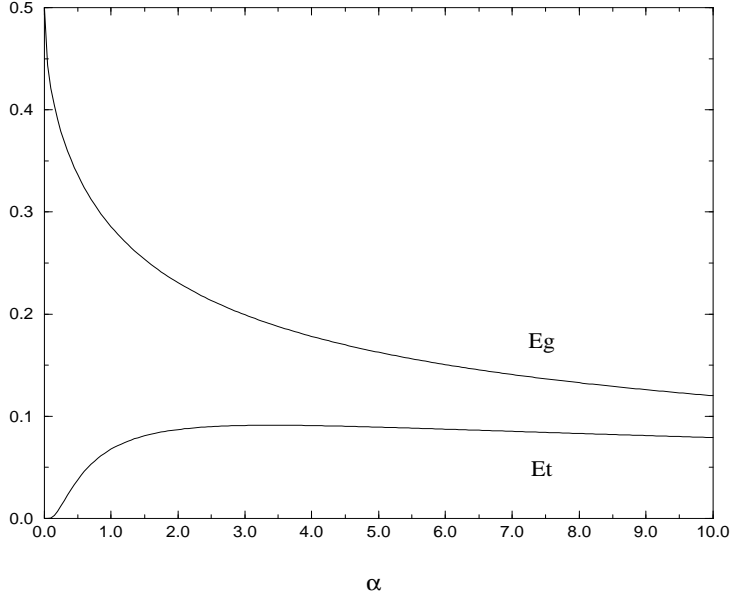}
\hspace*{1mm}}
\vspace*{-5mm}
\caption{Fig. 4: Asymptotic values of the two errors, i.e.
$\lim_{t\to\infty}E_{\rm g}$ and $\lim_{t\to\infty}E_{\rm t}$,
for Hebbian
learning (both on-line and batch), as functions of the relative
size
$\alpha=p/N$ of the training set.}
\end{figure}
\be
E_{\rm g}=\frac{1}{\pi}\arccos\left[\frac{R}{\sqrt{Q}}\right]
~~~~~~~~
E_{\rm t}=\frac{1}{2}-\frac{1}{2}\int\!Dy~
{\rm erf}\left[\frac{|y|R\plus \eta t/\alpha}{\sqrt{2(Q\minus
R^2)}}\right]
\label{eq:Hebberrors}
\ee
Asymptotically one finds $\lim_{t\to\infty} q=1$, and
\bd
\lim_{t\to \infty}E_{\rm g}=
\frac{1}{\pi}\arccos
\left[\frac{\sqrt{2\alpha}}{\sqrt{2\alpha+\pi}}\right]
~~~~~~~~
\lim_{t\to\infty}E_{\rm t}=
\frac{1}{2}-\frac{1}{2}\int\!Dy~
{\rm erf}\left[|y|\sqrt{\frac{\alpha}{\pi}}+
\frac{1}{\sqrt{2\alpha}}\right]
\ed
Both asymptotic errors are independent of $\kappa$, i.e.
on whether batch
or on-line learning is used.
These results are depicted in Fig. 4.

A final object to be calculated
is the student field distribution $P(x)$ itself, via
$P(x)=\int\!dy~P[x,y]=\int\!Dy~
P[x|y]$. This gives
\bd
P(x)
=\frac{e^{-\frac{1}{2}[x+\eta t/\alpha]^2/Q}}
{2\sqrt{2\pi Q}}
\left[1-
{\rm erf}
\left(\frac{R[x\plus\eta t/\alpha]}{\sqrt{2Q(Q\minus R^2)}}
\right)\right]
~~~~
\ed
\be
~~~~
+
\frac{e^{-\frac{1}{2}[x-\eta t/\alpha]^2/Q}}{2\sqrt{2\pi Q}}
\left[1+{\rm erf}\left(\frac{R[x\minus \eta
t/\alpha]}{\sqrt{2Q(Q\minus R^2)}}\right)\right]
\label{eq:Hebbstudentdist}
\ee

\subsection{Comparison with Exact Results and Simulations}

Only for the (simple) Hebbian rule can our dynamic order
parameters
in fact be
calculated  directly from the microscopic learning
rules, even for finite $\alpha$, which provides
an excellent benchmark for candidate general theories.
Exact evaluation
for on-line learning is found to give (Rae et al, 1998):
\be
R= R_0+\eta
t\sqrt{\frac{2}{\pi}}
~~~~~~~~
Q=Q_0+\eta t\left[\eta\plus 2 R_0\sqrt{\frac{2}{\pi}}\right]
+\eta^2 t^2\left[
\frac{1}{\alpha}\plus \frac{2}{\pi}\right]
\label{eq:exactQR}
\ee
\be
P[x|y]=\int\!\frac{d\hat{x}}{2\pi}~e^{-\frac{1}{2}
\hat{x}^2[Q-R^2]
+i\hat{x}[x-yR]+\frac{t}{\alpha}[e^{-i\eta\hat{x}\sgn(y)}-1]}
\label{eq:exactP}
\ee
Comparison with (\ref{eq:HebbQR},\ref{eq:HebbP}) shows
that our theory gives the exact expressions for $Q$ and $R$
(and thus for $E_{\rm g}$),
but an approximation for $P[x|y]$ (and
thus for $E_{\rm t}$) as soon as both $\alpha$ and $t$
are finite.
The most transparent comparison is made in terms of the
Fourier transform
$\hat{P}[k|y]=\int\!dx~e^{-ikx}P[x|y]$:
\be
\hat{P}[k|y]_{\rm exact}=
e^{-\frac{1}{2}k^2(Q-R^2)-ikRy+\frac{t}{\alpha}
[e^{-i\eta k\sgn(y)}-1]}
\label{eq:exactfourier}
\ee
\be
\hat{P}[k|y]_{\tiny\rm DRT}=
e^{-\frac{1}{2}k^2(Q-R^2)-ikRy+\frac{t}{\alpha}[-i\eta
k\sgn(y)]}
\label{eq:drtfourier}
\ee
One obtains
$\hat{P}[k|y]_{\tiny\rm DRT}$ by retaining
only the first two orders in the expansion of the
term $e^{-i\eta k\sgn(y)}$ in the exponent of
$\hat{P}[k|y]_{\rm exact}$.
The difference between the exponents of
(\ref{eq:exactfourier}) and (\ref{eq:drtfourier})
is in $\order(t)$ rather than $\order(t^2)$ terms. Since the
latter control the asymptotics,
exactness of our theory is (for any $\alpha$)
restored for $t\to\infty$; thus the asymptotic errors
shown in Fig. 4 are also exact.

\begin{figure}[t]
\centering
\vspace*{90mm}
%fullhebbsimu.eps
\hbox to \hsize{\hspace*{-1mm}\includegraphics{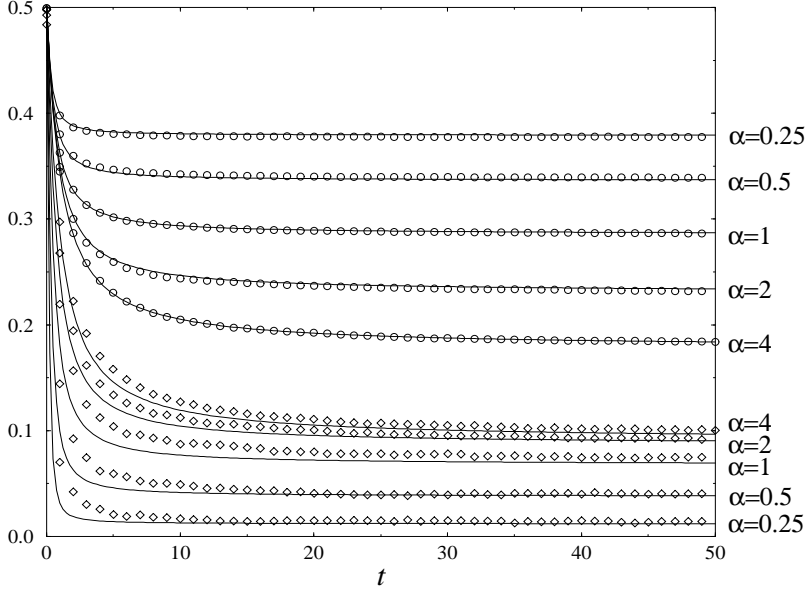}
\hspace*{1mm}}
\vspace*{-10mm}
\caption{Fig. 5: Comparison between simulation results for
on-line Hebbian
learning (with system size $N=10,000$) and dynamical
replica theory,
for $\alpha\in\{0.25,0.5,1.0,2.0,4.0\}$. Upper
five curves: generalization errors as functions of time.
Lower five
curves: training errors as functions of time.
Circles: simulation results for generalization errors;
diamonds: simulation results for training errors.
Solid lines: corresponding theoretical predictions.}
\end{figure}

We conclude: either replica symmetry must be broken
(RSB), or
our set of order parameters $\{Q,R,P\}$
does not yet obey closed deterministic and self-averaging
dynamical
equations
(otherwise our theory would have been exact, by
construction). The natural ways to try improve our
theory are thus either to construct RSB solutions of
the saddle-point equations,
or  to add observables to the order parameter
set,
such as the Green's function (\ref{eq:green}).
\clearpage

\begin{figure}[ht]
\centering
\vspace*{65mm}
\hbox to
\hsize{\hspace*{5mm}\includegraphics{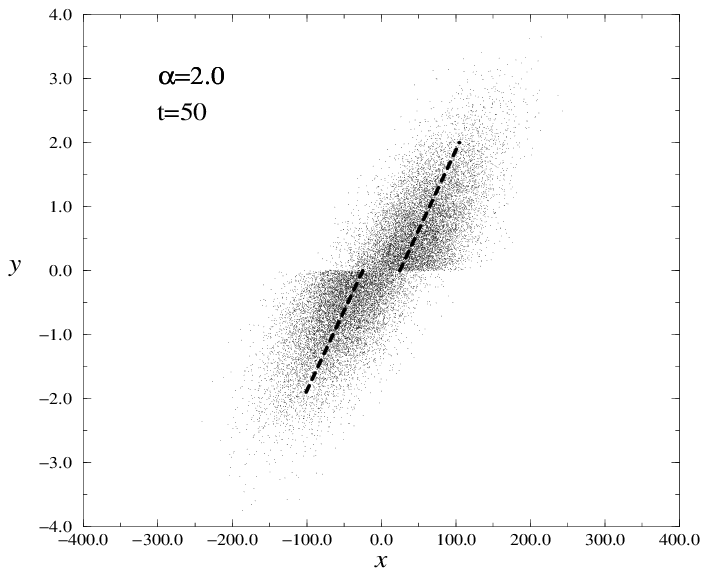}\hspace*{-5mm}}
\vspace*{55mm}
\hbox to
\hsize{\hspace*{5mm}\includegraphics{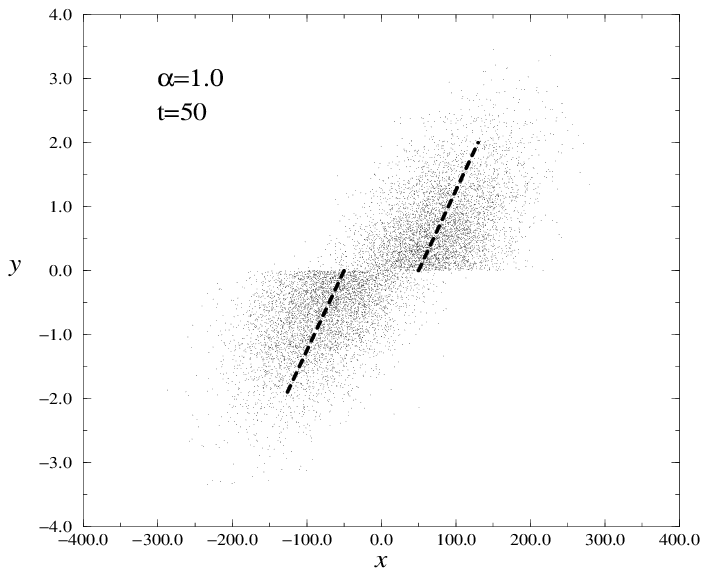}\hspace*{-5mm}}
\vspace*{55mm}
\hbox to
\hsize{\hspace*{5mm}\includegraphics{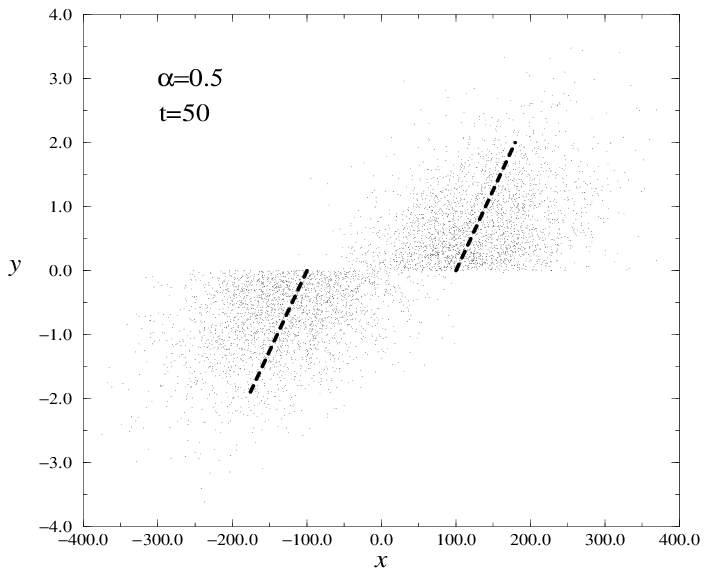}\hspace*{-5mm}}
\vspace*{-5mm}
\caption{Fig. 6: Comparison between simulation results for on-line
Hebbian
learning (system size $N=10,000$) and dynamical replica theory,
for $\alpha\in\{0.5,1.0,2.0\}$. Dots: local fields
$(x,y)=(\bJ\inn\bxi,\bB\inn\bxi)$ (calculated for questions in the
training set),
at time  $t=50$. Dashed lines: conditional average of student field
$x$ as a function of $y$, as predicted by the theory,
$\overline{x}(y)=Ry+(\eta t/\alpha)\sgn(y)$.}
\end{figure}

\begin{figure}[ht]
\centering
\vspace*{55mm}
\hbox to
\hsize{\hspace*{8mm}\includegraphics{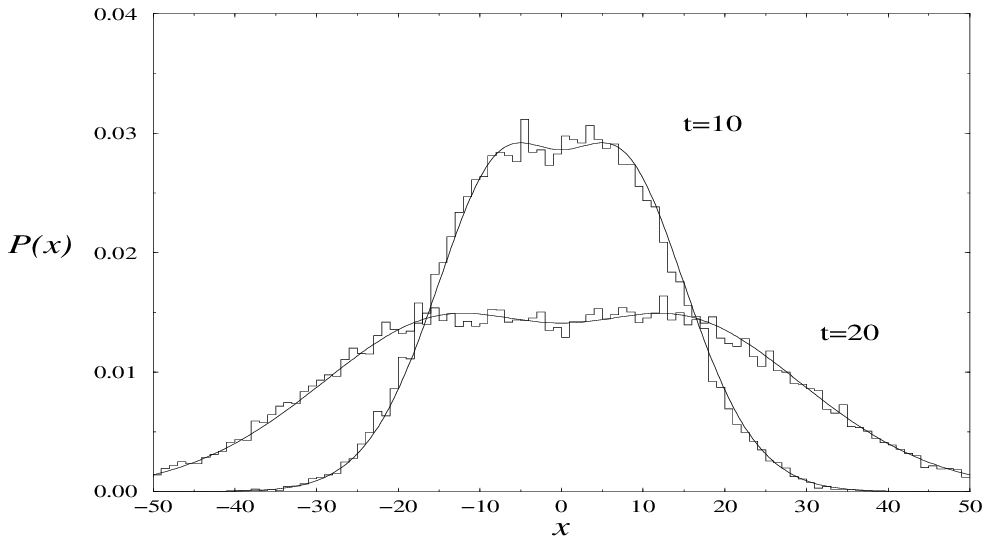}\hspace*{-8mm}}
\vspace*{52mm}
\hbox to
\hsize{\hspace*{8mm}\includegraphics{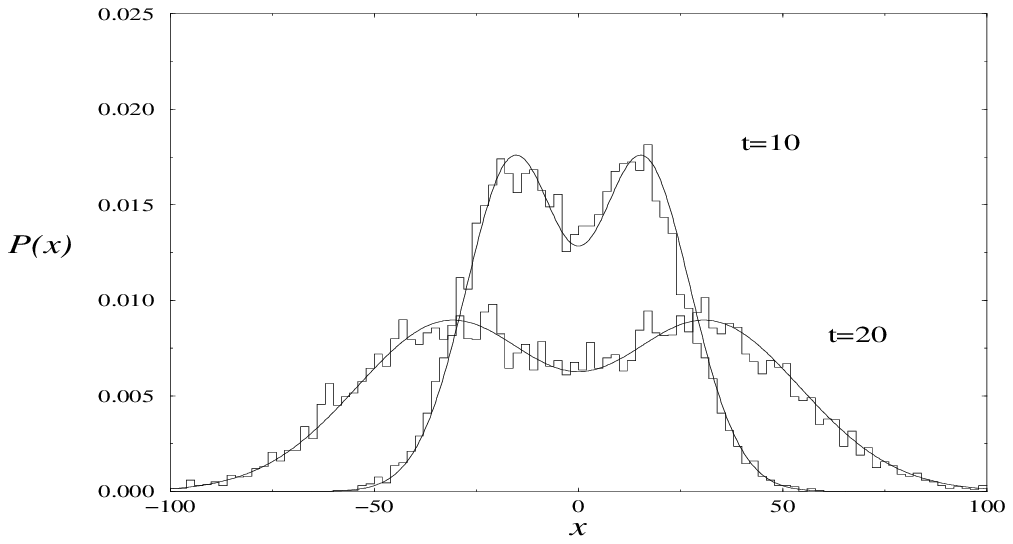}\hspace*{-8mm}}
\vspace*{52mm}
\hbox to
\hsize{\hspace*{8mm}\includegraphics{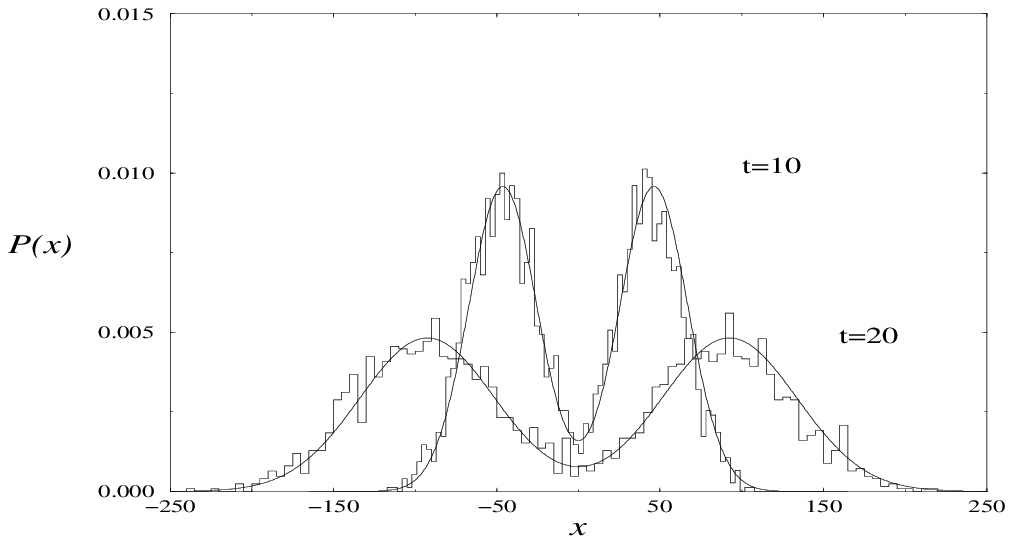}\hspace*{-8mm}}
\vspace*{-2mm}
\caption{Fig. 7: Simulations of Hebbian on-line learning  with $N=10,000$.
Histograms: student field distributions measured at $t=10$ and $t=20$.
Lines: theoretical predictions for student field distributions.
$\alpha=4$ (upper), $\alpha=1$ (middle), $\alpha=0.25$
(lower).}
\end{figure}
\clearpage

In Fig. 5 we compare the
predictions
for the generalization and training errors ({\ref{eq:Hebberrors})
of our theory with the results
obtained from numerical simulations for $N=10,000$
(initial state: $Q_0=1$, $R_0=0$; learning rate: $\eta=1$).
The curves for the generalization errors show full agreement between
theory and
experiment, as
guaranteed by (\ref{eq:exactQR}). However,
the transients of $E_{\rm t}$ show deviations, which become
more
pronounced with decreasing $\alpha$, but which vanish with
increasing time. This trend immediately
follows from (\ref{eq:exactfourier},\ref{eq:drtfourier}).

We also compare the theoretical predictions made for the
distribution $P[x|y]$ with the results of numerical simulations.
This
is done in Fig. 6, where we show the fields as
observed at time $t=50$ in simulations ($N=10,000$, $\eta=1$,
$R_0=0$,
$Q_0=1$) of on-line Hebbian learning, for three different
values of
$\alpha$. In the same figure  we draw (as dashed lines) the
theoretical prediction for the $y$-dependent average of the
conditional $x$-distribution $P[x|y]$:
\bd
\overline{x}(y)=Ry+\sgn(y)\eta t/\alpha
\ed
We observe that our expression (\ref{eq:HebbP}) for
$P[x|y]$
(a Gaussian distribution in $x$, with $y$-dependent
average given by
$\overline{x}(y)$ as given above) indeed captures the
qualitative features of the $(x,y)$ statistics. Clearly,
$P[x,y]$ is
itself not a joint Gaussian distribution.

Finally we compare the student field distribution
$P(x)$, as
observed in simulations
of on-line Hebbian learning ($N=10,000$, $\eta=1$, $R_0=0$,
$Q_0=1$)
with our prediction (\ref{eq:Hebbstudentdist}). The result
is
shown in Fig. 7, for $\alpha\in\{4,1,0.25\}$.
The agreement is again quite satisfactory, as could
have been expected.

\section{Discussion}

In this paper we have shown how the formalism of dynamical
replica
theory (e.g. Coolen et al, 1996)  can be used successfully
to build
a general theory with which to
predict the evolution of the relevant macroscopic performance
measures
for supervised (on-line and batch) learning in layered neural
networks
with randomly composed but restricted training sets
(i.e. for finite $\alpha=p/N$),
where the student fields are no longer described by
Gaussian
distributions, and where the more traditional and familiar
statistical mechanical
formalism consequently  breaks down.
 For simplicity and transparency
we have restricted ourselves to single-layer systems and
realizable tasks.
In our approach the joint field
distribution $P[x,y]$
for student and teacher fields is itself taken to be a
dynamical
order parameter, in addition to the more conventional
observables
$Q$ and $R$; from this order parameter set $\{Q,R,P\}$,
in turn,
immediately
follow the generalization error $E_{\rm g}$ and the training
error
$E_{\rm t}$. This then results, following the prescriptions of
dynamical
replica theory\footnote{The reason why replicas are inevitable
(unless we are willing to pay the price of having observables
with two
time arguments, and turn to path integrals) is
the necessity for finite $\alpha$
to average the macroscopic equations over all possible
realizations of the training
set.}, in a diffusion equation for $P[x,y]$, which we
have evaluated by making the replica-symmetric ansatz in the
saddle-point equations. This diffusion equation
is found to have Gaussian
solutions only for $\alpha\to\infty$; in the latter case we
indeed
recover correctly from our
theory the more familiar formalism of infinite training sets,
with
(in the $N\to\infty$ limit)
closed
equations for $Q$ and $R$ only. For finite $\alpha$ our theory
is by
construction exact
if for $N\to\infty$ the dynamical order parameters $\{Q,R,P\}$
obey closed,
deterministic
equations, which are self-averaging (i.e.
independent of the microscopic realization of the training set).
If this is not the case, our theory is an approximation.

We have worked out our general equations explicitly for the
special
case of Hebbian learning, where the existence of exact results,
derived directly from the microscopic equations
(even for finite $\alpha$), allows us to perform a critical
test of
our theory \footnote{Note that such exact results can only be
obtained for the relatively simple Hebbian rules, where
the dependence of the weight updates $\Delta\bJ(t)$ on the
current weights $\bJ(t)$ is trivial or even absent (a decay
term
at most),
whereas our present theory generates
macroscopic equations for arbitrary learning rules.}.
Here we find that our theory does produce
correct predictions for the observables $Q$, $R$ and
$E_{\rm g}$,
but an approximation for $P[x,y]$ and $E_{\rm t}$ if
$\alpha$ is finite
%%%%%%%%%%%%%%%%%% minor modification
(although the stationary state predicted is again correct).

The present study clearly represents only a first step,
and many
extensions, applications and generalizations can and
should be
made. To name but a few:
\\[1mm]
{\em (i) Application to Different Learning Rules}
\\[1mm]
So far our theory has only been applied to Hebbian
learning,
in view of its special status as a rigorous benchmark
(Rae et al, 1998). Further  application to non-Hebbian
learning rules, subsequently to be tested via
simulations, is clearly called for.  For rules of the type
(\ref{eq:localgauss}) the saddle-point equations are still
simple; in general one will have to solve functional
saddle-point equations at each instance of time, which is a
non-trivial numerical exercise.
\\[1mm]
{\em (ii) Application to Multi-Layer Networks}
\\[1mm]
Our theory generalizes in a natural way to multi-layer
networks,
provided the number of hidden neurons remains finite.
However,
as in the infinite training set formalism, the number of
observables (and thus the number of saddle-point equations)
will
increase significantly.
\\[1mm]
{\em (iii) Further analysis of saddle-point equations}
\\[1mm]
We still have to determine the uniqueness or otherwise
of the solution of our functional saddle-point equation.
More ambitious, but not ruled out, are our current attempts
to solve the functional saddle-point equation explicitly.
If this is impossible, one further alternative to
numerical solution would be a variational approach
(where upon choosing a restricted parametrized family of
functions,
 functional extremization is replaced by ordinary
extremization).
\\[1mm]
{\em (iv) Replica Symmetry Breaking}
\\[1mm]
The observed deviations between the
present theory and the exact calculation for the Hebb rule could
indicate RSB (although this is not likely, in view of the
asymptotic exactness of our RS equations). In the usual manner
one can calculate an equation for the
AT instability, which would in our problem define a surface in
order parameter space, and determine the onset of RSB.
\\[1mm]
{\em (v) Systematic Improvement via Higher Order Observables}
\\[1mm]
Dynamical replica theory allows for systematic improvement. By
adding new
observables to the order parameter set (which cannot be expressed
in terms of those already present) the theory will,  by
construction,  become more accurate. A natural candidate
for being added to the set $\{Q,R,P\}$ is the Green's function
$\cA[x,y;x^\prime,y^\prime]$ (\ref{eq:green}). This would change
our problem to closure of a
dynamic equation for $\cA$, which would involve a higher order
Green's function.
\\[1mm]
{\em (vi) Generalization to Noisy Teachers}
\\[1mm]
Last but not least, one can generalize our theory to the case of
noisy teachers. This is a straightforward although tedious
exercise, involving a field distribution of the form $P[x,y,z]$
(describing student fields, fields of the `perfect' teacher, and
fields of the `noisy' teacher). It will, however, allow us to
describe over-fitting phenomena in terms of macroscopic dynamic
equations.

\vspace*{5mm}
\noindent
{\bf Acknowledgement:}\\ DS acknowledges support by EPSRC
Grant GR/L52093.

\section*{References}

\refce
Barber D., Saad D. and Sollich P. (1996),
{\em Europhys. Lett.} {\bf 34}, 151

\refce
Biehl M. and Schwarze H. (1992),
{\em Europhys. Lett.} {\bf 20}, 733

\refce
Biehl M. and Schwarze H. (1995),
{\em J. Phys. A: Math. Gen.} {\bf 28}, 643

\refce
Coolen A.C.C., Laughton S.N. and Sherrington D. (1996),
{\em Phys. Rev. B} {\bf 53}, 8184

\refce
Coolen, A.C.C and Saad D. (1998), in preparation.

\refce
Horner H. (1992a), {\em Z. Phys. B} {\bf 86}, 291

\refce
Horner H. (1992b), {\em Z. Phys. B} {\bf 87}, 371

\refce
Kinouchi O. and Caticha N. (1992),
{\em J. Phys. A: Math. Gen.} {\bf 25}, 6243

\refce
Kinzel W. and Rujan P. (1990),
{\em Europhys. Lett.} {\bf 13}, 473

\refce
Mace C.W.H. and Coolen A.C.C (1998a), {\em Statistics and Computing},
 in press

\refce
Mace C.W.H. and Coolen A.C.C (1998b), in preparation

\refce
M\'{e}zard M., Parisi G. and Virasoro M.A. (1987), {\em Spin-Glass
Theory and Beyond} (Singapore: World Scientific)

\refce
Rae H.C., Sollich P. and Coolen A.C.C. (1998), in preparation

\refce
Saad D. and Coolen, A.C.C (1998), in preparation

\refce
Saad D. and Solla S. (1995),
{\em Phys. Rev. Lett.} {\bf 74}, 4337

\refce
Sollich P. and Barber D. (1997), to be published in Proc. NIPS*97

\end{document}